\documentclass[%
10pt,
 twocolumn,
amsmath,amssymb,
 aps,
prb,
longbibliography,
superscriptaddress
]{revtex4-2}

\usepackage{natbib,hyperref}
\usepackage{blindtext}
\usepackage{graphicx}
\usepackage{dcolumn}
\usepackage{svg}
\usepackage{bm}
\usepackage{mathtools}
\usepackage{amssymb}
\usepackage{amsmath}
\usepackage{amsthm}
\usepackage{graphicx}
\usepackage{listings}
\usepackage{epsfig}
\usepackage{amsfonts}
\usepackage{bbold}
\usepackage{xcolor}
\usepackage{orcidlink}
\usepackage{multirow} 
\usepackage{footnote}
\usepackage{blkarray}
\usepackage[]{appendix}
\usepackage{footnote}
\makesavenoteenv{table} 
\usepackage[makeroom]{cancel}
\makeatletter

\def\cantox@vector#1#2#3#4#5#6#7#8{%
  \dimen@.5\p@
  \setbox\z@\vbox{\boxmaxdepth.5\p@
   \hbox{\kern-1.2\p@\kern#1\dimen@$#7{#8}\m@th$}}%
  \ifx\canto@fil\hidewidth  \wd\z@\z@ \else \kern-#6\unitlength \fi
  \ooalign{%
    \canto@fil$\m@th \CancelColor
    \vcenter{\hbox{\dimen@#6\unitlength \kern\dimen@
      \multiply\dimen@#4\divide\dimen@#3 \vrule\@depth\dimen@\@width\z@
      \vector(#3,-#4){#5}%
    }}_{\raise-#2\dimen@\copy\z@\kern-\scriptspace}$%
    \canto@fil \cr
    \hfil \box\@tempboxa \kern\wd\z@ \hfil \cr}}
\def\bcancelto#1#2{\let\canto@vector\cantox@vector\cancelto{#1}{#2}}
\makeatother
\usepackage[]{hyperref}
\usepackage{soul} 

\newcommand\ket [1]{\left| #1 \right>}
\newcommand\bra [1]{\left< #1 \right|}

\newcommand{\corr}[1] {{\color{black}{#1}}\color{black}}
\newcommand{\cor}[1] {{\color{black}{#1}}\color{black}}
\renewcommand{\eth}[0]{Department of Materials, ETH Z{\"u}rich, 8093 Z{\"u}rich, Switzerland}

\newcommand{\expSTMESR}{~\cite{Baumann_Paul_science_2015,
Natterer_Yang_nature_2017,Choi2017a,Willke_Paul_sciadv_2018,Yang_Bae_prl_2017,
Willke_Bae_science_2018,Y_Bae_advanced_science_2018,Willke_Singha_nanolett_2019,
Willke_Yang_natphys_2019,Yang_Paul_prl_2019,Seifert_Kovarik_pr_2020,
Seifert_Kovarik_eabc_2020,
Weerdenburg_Steinbrecher_2020,Steinbrecher_Weerdenburg_2020,Jinkyung_Hyperfine_2022,zhang_electron_2022,kovarik_electron_2022,kot2023electric,zhang_influence_2023,Spin_torque-driven_Kovarik_2024,czap2025magneticresonanceimagingsingle, zhang2024electricfieldcontrolexchange,Huang2025,greule2025spinelectriccontrolindividualmolecules}}

\begin{document}

\title{Efficient driving of a spin-qubit using single-atom magnets}

\author{Jose Reina-G\'alvez\,\orcidlink{0000-0001-7587-056X }}
\email{jose.reina-galvez@uni-konstanz.de}
\affiliation{Department of Physics, University of Konstanz, D-78457 Konstanz, Germany}
\affiliation{%
Center for Quantum Nanoscience, Institute for Basic Science, Seoul 03760, Republic of Korea
}%
\affiliation{%
Ewha Womans University, Seoul 03760, Republic of Korea
}%

\author{Hoang-Anh Le\,\orcidlink{0000-0002-1668-8984}}
\thanks{These authors contributed equally: Jose Reina-G\'alvez, Hoang-Anh Le}
\affiliation{%
Center for Quantum Nanoscience, Institute for Basic Science, Seoul 03760, Republic of Korea
}%
\affiliation{%
Ewha Womans University, Seoul 03760, Republic of Korea
}%

\author{Hong Thi Bui}
\thanks{Currently at: \eth}
\affiliation{%
Center for Quantum Nanoscience, Institute for Basic Science, Seoul 03760, Republic of Korea
}%
\affiliation{Department of Physics, Ewha Womans University, Seoul 03760,  Republic of Korea}

\author{Soo-hyon Phark\,\orcidlink{0000-0002-0541-5083}}
\affiliation{%
Center for Quantum Nanoscience, Institute for Basic Science, Seoul 03760, Republic of Korea
}%
\affiliation{%
Ewha Womans University, Seoul 03760, Republic of Korea
}%

\author{Nicol\'as Lorente\,\orcidlink{0000-0003-0952-8031}}
\affiliation{Centro de F{\'{i}}sica de Materiales
        CFM/MPC (CSIC-UPV/EHU),  20018 Donostia-San Sebasti\'an, Spain}
\affiliation{Donostia International Physics Center (DIPC),  20018 Donostia-San Sebasti\'an, Spain}    

\author{Christoph Wolf\,\orcidlink{0000-0002-9340-9782}}
\email{wolf.christoph@qns.science}
\affiliation{%
Center for Quantum Nanoscience, Institute for Basic Science, Seoul 03760, Republic of Korea
}%
\affiliation{%
Ewha Womans University, Seoul 03760, Republic of Korea
}%

\date{\today}

\begin{abstract}
The realization of electron-spin resonance at the single-atom level using scanning tunneling microscopy has opened new avenues for coherent quantum sensing and quantum state manipulation at the ultimate size limit. This allows to build many-body Hamiltonians and the study of their complex physical behavior. Recently, a novel qubit platform has emerged from this field, raising questions about the driving mechanism from single-atom magnets. In this work, we demonstrate how single-atom magnets can be used to drive a nearby single spin qubit efficiently. {\cor{ We show that the modulation of exchange coupling is the primary driving force, which successfully reproduces Rabi rates in the tens of MHz range, consistent with experimental data, }} while also addressing critical aspects related to the optimization of experimental parameters.
\end{abstract}


\maketitle

{\emph{Introduction}--} Since the first demonstration of electron-spin resonance (ESR) on the level of individual atoms using a scanning tunneling microscope (STM), the combination of ESR and STM has been expanded to the study of single atoms, pairs, and molecules\expSTMESR.
Utilizing pulsed ESR techniques has also been used to coherently control the quantum state of single~\cite{yang_coherent_2019, Wang2023UniversalSurface} and multiple electrons on surfaces~\cite{Wang2023AnPlatform}. At the same time, the mechanism that allows to drive ESR in an STM using time-varying electric fields has remained somewhat elusive. 

Originally, it was proposed that a modulation of the crystal field by the applied electrical field is responsible for creating the necessary transition matrix elements~\cite{Baumann_Paul_science_2015}. Then, it was suggested that the applied electrical field modulates the coupling between the magnetic STM tip apex and the spin~\cite{Lado_Ferron_prb_2017} {\corr{or changes the $g$-factor of the adatom}}~\cite{Ferron2019Anisotropy}.
And recently, it was suggested that the AC magnetic field component of the electro-magnetic field supplied by an antenna  in the STM junction region could serve as a driving field~\cite{J_Cuevas_C_Ast_ESR_theory_2024}.
Some of us showed that an {\cor{AC voltage}} is able to reduce or increase the tunneling barrier, which modulates the coupling between the tip and the spin~\cite{J_Reina_Galvez_2019,J_Reina_Galvez_2021, J_Reina_Galvez_2023,eric_unraveling_2025}. This modulation in presence of a polarized electrode leads to a time dependent exchange field which drives ESR {\cor{when on resonance with a spin transition}}~\cite{Braun_Konig_prb_2004,Rabi_paper}.
In all cases the model incorporated some influence from the STM tip, which, in typical STM fashion, can be expected to have a highly localized interaction with the spin on the surface. All of the previous approaches were challenged when it was discovered that a spin can be coherently driven \textit{without} being directly under the tip apex, {\corr{by placing a magnetic atom nearby~\cite{Wang2023AnPlatform, Phark2023Electric-Field-DrivenMagnet, Phark2023Double-ResonanceSurface, Bui2024All-electricalSpinb}}}. 

A well-demonstrated system is the engineered Fe-Ti pair~\cite{Phark2023Electric-Field-DrivenMagnet}. In this system, the Ti spin state can be driven either by the tip or by the nearby Fe atom. Here, the Fe atom can be approximated as a single-atom magnet since it possesses an anisotropy barrier against spin reversal, and the lifetime of its spin state is $\sim10^5$ times longer than the coherence time of the Ti spin~\cite{Paul_Yang_natphys_2017}. Using two Ti spins, where one spin played the role of the sensor while the other formed an atomic pair with an Fe atom, it was also shown that the driving mechanism works when the controlled spin is outside the tunnel junction~\cite{Wang2023_universal, Phark2023Double-ResonanceSurface, Wang2023AnPlatform, Bui2024All-electricalSpinb}. 
This experimental observation is not compatible with the assumption of an exchange modulation between the tip apex and the surface spin which has a range of a few \r{A}. However, the electric field that emanates from the wider tip body {\corr{or a nearby antenna covers}} an area of possibly several {\corr{mm}}$^2$, {\corr{which allows it}} to drive {\corr{spins outside of the tunnel junction}}~\cite{Wang2023_universal, Phark2023Double-ResonanceSurface, Wang2023AnPlatform, Bui2024All-electricalSpinb}. {\corr{This field could modulate the $g$-factor of the surface spins, the on-site crystal-field anisotropy terms, or the exchange coupling between Ti and Fe.}} {\corr{However, the absence of any measurable effect of the electric field on the magnetic moment observed experimentally rules out the $g$-factor modulation as the main driving mechanism~\cite{zhang2024electricfieldcontrolexchange}, leaving two remaining possibilities.}}

In this work, we show that although crystal field modulation can drive a spin, the resulting Rabi rates are approximately an order of magnitude smaller for the Fe-Ti system, where the most reliable experimental data is available. To alleviate this discrepancy between experiment and model, we suggest a driving mechanism that is {\corr{entirely}} based on electric field modulation of the exchange coupling between the Ti and Fe. We will demonstrate that this approach leads to Rabi rates {\cor{ in the range of $10 - 30 \text{ MHz}$,}} compatible with experimental data and provides a better understanding of how spins on surfaces are driven by electric fields in ESR-STM.

\begin{figure}
\centering
\includegraphics[width=\linewidth]{ 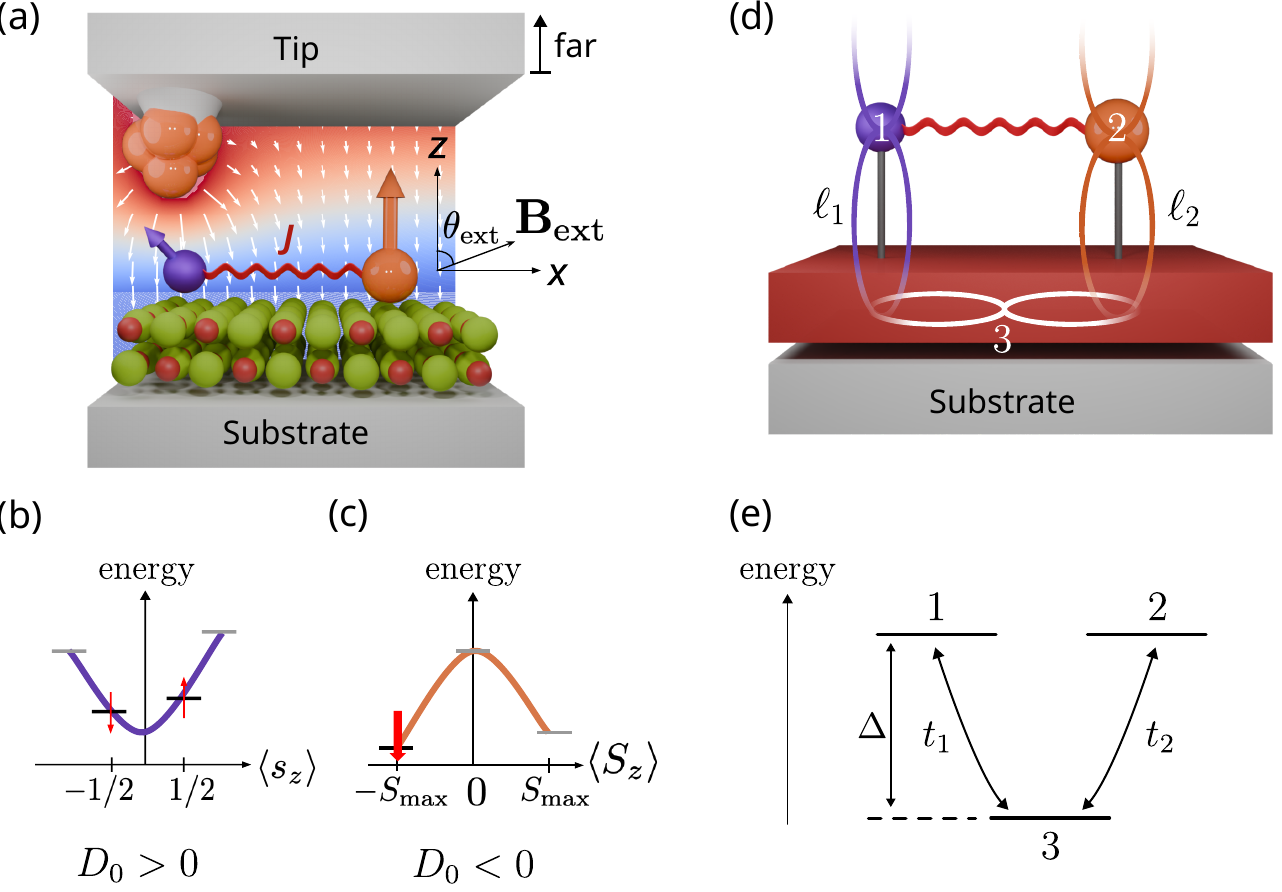}
\caption{{Spin-qubit system:} (a) Front view of two atoms subject to an electric field in the STM junction.  External magnetic field direction as indicated. The atom-pair is exchange coupled with coupling strength $J$. (b) Spin qubit, i.e. a spin$-1/2$ without magnetic stability $D_0>0$. (c) Single-atom magnet $S>1/2$ with an energy barrier $D_0<0$. (d) {\cor{ Substrate-mediated superexchange of an atomic pair. Atom $1$ and $2$ are connected via the substrate wave-function represented by a single effective orbital $3$. $\ell_{1(2)}$ is the vertical distance from orbital 1(2) to the effective orbital 3. (e) $t_{1(2)}$ represents the hopping amplitude from orbital 1(2) to effective orbital 3 and $\Delta$ the site potential difference.}}}
\label{fig:fig1}
\end{figure}

{\emph{Exchange-coupled spin pair}--} We first focus on a generic pair of two spins coupled on a surface via an isotropic Heisenberg interaction and subjected to an external magnetic field as indicated in Fig.~\ref{fig:fig1}(a). {\cor{Throughout this work we will discard the driving field provided by the exchange with the magnetic tip entirely, i.e. we consider the limit where the tip is far, and focus on the Heisenberg coupling between the two spins parameterized by $J$.}}
One spin is either a spin $s=1/2$ which does not have any anisotropy or a system with a ground state doublet not separated by an anisotropy barrier (i.e., $D_0>0$), see Fig.~\ref{fig:fig1}(b). This renders it an ideal candidate for a spin-qubit system and, therefore, it can be efficiently driven using ESR-STM~\cite{yang_coherent_2019, reale2023erbium, Spin_torque-driven_Kovarik_2024}. The other spin in the pair is a {\cor{single-atom magnet with a large anisotropy barrier}}~\cite{Donati2016atommagnet}, see Fig.~\ref{fig:fig1}(c). The simplest effective spin Hamiltonian to describe it is of the form $H_{D_0}=D_0 S_{z}^2$, where $D_0<0$ is the uniaxial anisotropy parameter, $S_{z}$ is the spin projection of $\bm{S}$ along quantization axis $z$, and $S\geq 1$. 
To distinguish these two spins, we will denote the spin-qubit with lower case $\bm{s}$ and the single-atom magnet with capital $\bm{S}$ throughout the remainder of the text. The Zeeman Hamiltonian for each spin reads $H_{\rm Z}=g\mu_{\rm B} \bm{B}_\mathrm{ext}\cdot\bm{s}$ (or $\bm{S}$), with $\mu_{\rm B}$ the Bohr magneton and $\bm{B}_\mathrm{ext}$ the external static magnetic field defined in polar coordinates $(B_{\mathrm{ext}},\theta_\text{ext})$.

The spin pair has a Hilbert space of the form $\mathcal{H}=\bm{s} \otimes \bm{S}$. In the following, we will employ the notations $\ket{\uparrow}$ and $\ket{\downarrow}$ for spin-qubit states, and $\ket{\Uparrow}$ and $\ket{\Downarrow}$ for single-atom magnet states. In this context, $\ket{\uparrow, \Uparrow}\leftrightarrow \ket{\downarrow, \Uparrow}$ corresponds to a spin-flip on the spin-qubit while the single-atom magnet remains unaffected.
Conversely, $\ket{\downarrow, \Uparrow}\leftrightarrow \ket{\downarrow, \Downarrow}$ would indicate the single-atom magnet has reversed its spin. A physical system corresponding to this spin pair consists of an iron (Fe, $S=2$) as single-atom magnet and a titanium (Ti, $s=1/2$) atom as spin-qubit~\cite{Phark2023Electric-Field-DrivenMagnet,Wang2023AnPlatform}. We will discuss the specific parameters for this system at the end. 
The static Hamiltonian of this spin pair is
\begin{equation}
H_{0} = g\mu_\mathrm{B}  \bm{s} \cdot  \bm{B}_{\text{ext}} + g\mu_\mathrm{B} \bm{S} \cdot \bm{B}_{\text{ext}} + D_0 S_z^2 - J_0  \bm{S} \cdot \bm{s} ,
\label{eq:stat_hamil}
\end{equation}
where $J_0$ indicates the static exchange coupling of the two spins. The underlying physical mechanism in the systems considered here is substrate-mediated super-exchange~\cite{Strandberg2010} which can be described by an effective three-orbital model, as illustrated in Fig.~\ref{fig:fig1}(d). In the limit of low-energy physics where the on-site repulsion $U$ is the dominant energy scale, $J_0$ can be represented as~\cite{Furuya2021,Koch:1018555}:
\begin{equation}
J_0 = \frac{4 \vert t_1 t_2 \vert^2}{(U + \Delta)^2} \left( \frac{1}{U} + \frac{1}{U + \Delta} \right),
\label{eq:J0}
\end{equation}
where $t_{1(2)}$ is the hopping amplitude for orbital $1(2)$ and $\Delta$ is the on-site energy difference between the spin centers 1 and 2 and the mediating orbital 3, depicted in Fig.~\ref{fig:fig1}(e). A detailed derivation is presented in the subsection A of~\cite{SM1}.



{\emph{Modulation of the anisotropy and exchange term}--} The last two terms in the Eq.~(\ref{eq:stat_hamil}) show two venues for modulation, $D_0$ and $J_0$. We note that it has been demonstrated that that \emph{DC} electric control of $D_0$ and $J_0$ allows for full control of a spin-1/2 system under suitable conditions~\cite{Switzer_PhysRevA_2021, Switzer_2022}. In the following, we want to show that an \emph{AC} modulation will exhibit the same physics and will allow to directly drive $\bf{s}$.

We first consider that an applied electrical field can modulate the crystal field of the adatom via the piezoelectric effect~\cite{GeorgePRL2012}.
Assuming that the adatom is not entirely neutral or has a non-vanishing (induced) dipole moment, the electric field could either pull the atom away from the substrate or push it closer. This process has to be adiabatic {\cor{\cite{Lado_Ferron_prb_2017, Rodriguez2025} }}, which is a reasonable assumption since the resonances (phonon modes) of the substrate-adatom are typically in the THz range while ESR operates in the regime of a few tens of GHz. As a result the AC voltage $\cos{(\omega_0 t + \varphi)}$ modulates the anisotropy in-phase, so the resulting time-dependent Hamiltonian is given by
\begin{equation}
\begin{aligned}
H_{D_1} (t) =  D_1 \cos (\omega_{ij} t) S_z^2,
\label{eq:D_mod}
\end{aligned}
\end{equation}
where we have set $\varphi = 0$, which corresponds to a fixed axis for the rotation. Varying $\varphi$ allows for full control of quantum state on the Bloch sphere~\cite{SM4}. To satisfy the resonance condition, the modulation frequency equals transition frequency between $\ket{i} = \ket{\uparrow, \Uparrow}$ and $\ket{j} = \ket{\downarrow, \Uparrow}$, in which the single-atom magnet remains in its spin-up state, i.e. $\omega_0 = \omega_{ij}$ (the spin-down state provides similar physics).

Second, if the exchange between adatoms is modulated by an electric field ~\cite{Troiani2019ManipulationAnisotropy},  a modulation of $J_0$ - given certain other geometric conditions - could provide a driving field with sufficient magnitude to drive ESR on the Ti atom given by
\begin{equation}
\begin{aligned}
H_{J_1} (t) =  -J_1 \cos (\omega_{ij} t ) \bm{S} \cdot \bm{s} .
\label{eq:J_mod}
\end{aligned}
\end{equation}

Using Eq.~\ref{eq:J0} we assume the modulation of exchange coupling via electric field is induced by the variation of on-site potential $\Delta$, which leads to $J_1\propto J_0 V_{AC}$. 
We note that the linear relation holds as long as the electric energy is much smaller than on-site repulsion $U$, i.e.  $ e (\ell_1 + \ell_2) |\mathbf{E}| /U \ll 1$, where $\ell_{1(2)}$ is the distance from orbital 1(2) to effective orbital 3 and $e$ is the electron charge. This is valid for the typically small amplitudes of the AC voltage where the change in on-site energies remain negligible compared to the on-site repulsion energy in the doubly occupied orbitals, see subsection B in~\cite{SM1}.
Finally, we emphasize that, although we are modulating the anisotropy of $\bm{S}$ and exchange coupling between the two spins, we are ultimately interested in the time evolution of $\bm{s}$.

\begin{figure}
    \centering    
\includegraphics[width=1.0\linewidth]{  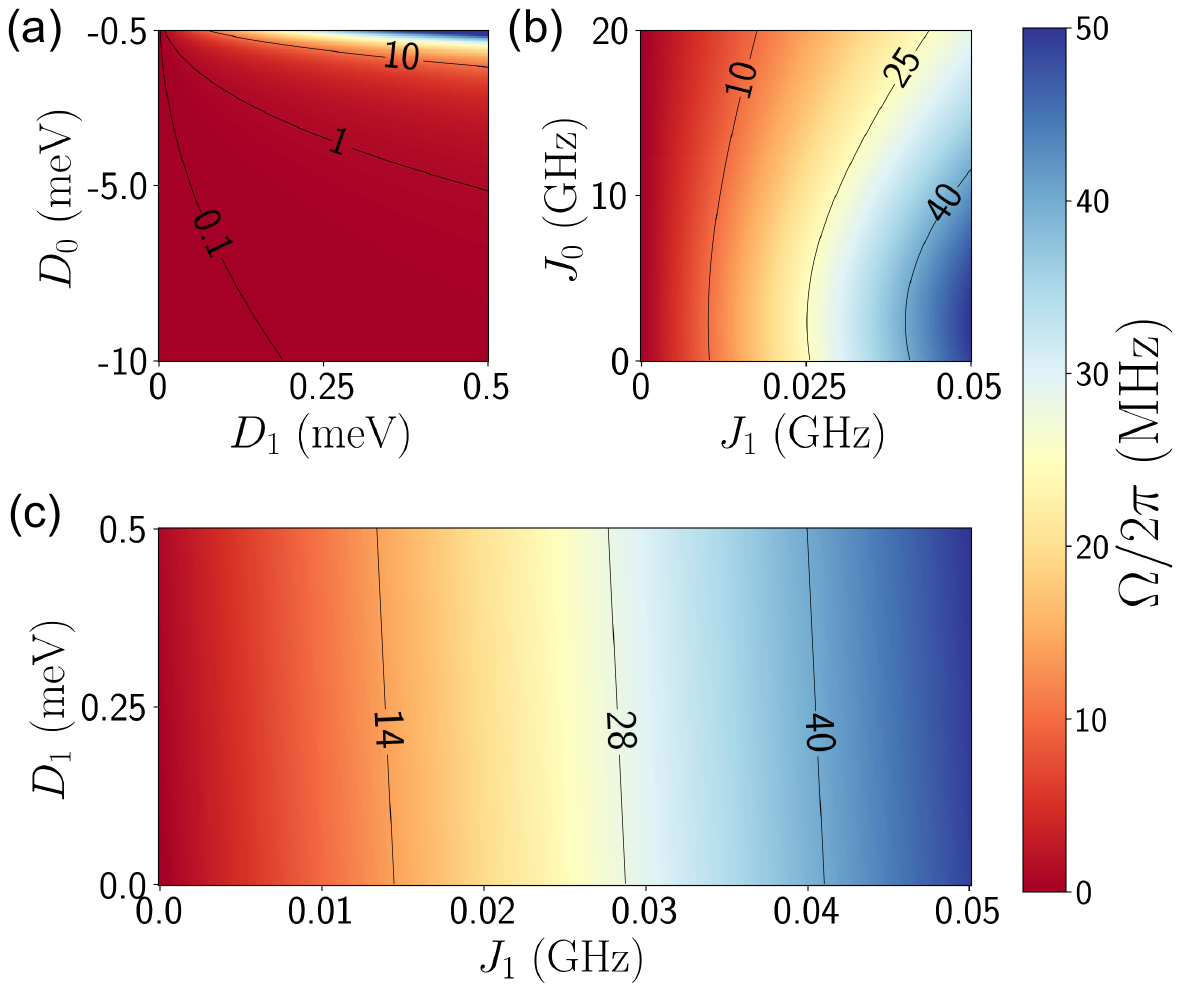}
    \caption{{Achievable Rabi rates from $D$, $J$ and $DJ$ modulation models:} (a), (b) Dependence of the Rabi rate by separate tuning of $D$ and $J$ on keeping the respective other parameters fixed. (c) Dependence of the Rabi rate on $J_1$ and $D_1$ while choosing $D_0$ and $J_0$ values compatible with the physical Fe-Ti system. 
    Fixed parameters are as follows: (a) $J_0 = 5 \textrm{ GHz}$ and $J_1 = 0 \textrm{ GHz}$; (b) $D_0 = -5$ meV and $D_1 = 0 \textrm{ meV}$; and (c) $D_0 = -5 $ meV and $J_0 = 5 \text{ GHz}$. Parameters related to the external magnetic field are: $B_\text{ext} = 0.9 \text{ T}$, $\theta_\text{ext}= 80^{\circ}$. We use isotropic $g$-tensor $g=2$.}
    \label{fig:mod_D1_J1}
\end{figure}

{\emph{Qualitative estimate of Rabi rates}--}  We now aim to qualitatively predict Rabi rates of $\bm{s}$. We solve the full Hamiltonian $H_0+H_{D_1}(t)+H_{J_1}(t)$ to obtain the time-evolution of both $\bm{S}$ and $\bm{s}$. 
The Rabi rate, $\Omega$, can be determined via the off-diagonal terms of time-dependent Hamiltonian, which read
\begin{equation}
  \hbar  \Omega = \bra{\uparrow, \Uparrow} (D_1 S_z^2 -J_1 \bm{S} \cdot \bm{s}) \ket{\downarrow, \Uparrow}.
    \label{eq:H_mod}
\end{equation}

We first consider each driving contribution in Eq.~(\ref{eq:H_mod}) separately, namely $D$-modulation and $J$-modulation, for reasonable direction and magnitude of a magnetic field~\cite{Phark2023Electric-Field-DrivenMagnet}.
Figure~\ref{fig:mod_D1_J1}(a) shows the Rabi rate solely from $D$-modulation. To maximize the Rabi rate, the ratio $D_1/D_0$ has to be maximized, i.e. the static contribution $D_0$ should be small relative to the modulation amplitude $D_1$. However, in our model the limit of $D_0\rightarrow 0$ is not valid as the system would not result in a magnet {\cor {with large magnetic anisotropy energy}} and in any physical system, $D_0$ is given by the material and adsorption site. 
The result associated with the $J$-modulation in Fig.~\ref{fig:mod_D1_J1}(b) shows a different trend. Here, a relatively small ratio $J_1/J_0$ is able to achieve Rabi rate values significantly larger than the ones obtained by the $D$-modulation. This is further supported by Fig.~\ref{fig:mod_D1_J1}(c), where we combined both $DJ$-modulations for a given $J_0$ and $D_0$, since the Rabi rate remains nearly unchanged along $D_1$ direction while it changes dramatically along $J_1$ direction. Figure~\ref{fig:mod_D1_J1}(c) implies that, in the context of $J$-modulation, a magnitude able to explain the experimental data can be easily achieved or exceeded. Whilst each individual spatial component of the exchange modulation can drive the spin, the direction of the modulation that aligns with the anisotropy axis of the single atom magnet will dominate as we analyze in~\cite{SM3}.

{\emph{Quantitative estimate of Rabi rates}--} We now apply our model to a physical system. To mimic the experiment we will use parameters that are typical for a Fe-Ti pair on 2 {\corr{monolayers}} of MgO on a silver substrate. Fe is a well characterized single-atom magnet with $D_0=-4.7$ meV~\cite{Paul_Yang_natphys_2017}. For a small distance, the magnitude of $J_0$ depends exponentially on the separation between the two atoms on the surface~\cite{Choi2017a}. In Ref.~\cite{Phark2023Electric-Field-DrivenMagnet} exchange coupling values of $J_0=1.1\pm 0.1$ GHz and $6.8\pm0.5$ GHz were measured for $d=0.72$ nm and $0.59$ nm, respectively. These couplings were extracted from the splitting of the resonance frequencies considering the influence of the external magnetic field angle $\theta_\mathrm{ext}$, which effectively modifies the net exchange coupling between atoms.
In the following, we will apply our model to a pair of $d=0.59$ nm.

Regarding the driving contributions, we first assess the $D_1$ modulation of Fe on MgO/Ag. For details of the model, including calculations and experimental data, we refer to Ref.~\cite{Seifert2020LongitudinalMicroscope}. Here, it should suffice to calculate the modulation $D_1$ as a function of the displacement of the Fe adatom, i.e., $D_1(z)$, for which we performed a series of multiplet calculations using a multiorbital Hubbard model~\cite{Wolf_C_and_Delgado_F_2020}. Using typical RF field strengths of 0.1 V/nm, we expect a displacement of at most $0.05$ pm, as shown in~\cite{SM5}. This leads to a maximal variation of $D_1=0.41\; \mu\text{eV } (\approx 100 \text{ MHz})$, 
{\cor{contributing much less than $0.1 \text{ MHz}$ to the Rabi rates, according to the result shown in Fig.~\ref{fig:mod_D1_J1}(a).}}
We note that in this displacement range, the relation between $D$ and $z$ is strictly linear with a proportionality of $0.107$ meV/nm.


{\cor{We have confirmed that the magnitude of $J_1$ required to achieve the experimental Rabi rates are reasonable in this system. Based on experimental data for the Rabi rate of Fe-Ti pairs, the magnitude of $J_1/\Delta E$ should be of the order of 20 MHz/0.1 V/nm~\cite{Bui2024All-electricalSpinb}}}. This is compatible with results from our DFT calculations on a magnetically coupled Fe-Fe and Fe-Ti pair adsorbed on the oxygen-top site of a MgO/Ag slab. More details of DFT calculations can be found in~\cite{SM5}. We note that on-surface exchange for Fe-Ti and Fe-Fe pairs might be highly anisotropic, considering the ground-state wave function of, for example, Ti~\cite{Jinkyung_Hyperfine_2022, Farinacci2022}. Typical values for the field modulation of $J$ reported in literature for other systems are approximately 6$-$40 times larger than those required for Fe-Ti pairs, and can exceed them by several orders of magnitude for molecular magnets. A comparison for various magnetic systems is provided in~\cite{SM2}. Therefore, it is reasonable to assume that the order of $J_1$ in Fig.~\ref{fig:mod_D1_J1}, from 5$-$50 MHz, is very plausible in our model. 
We also emphasize that, considering the results shown in Figs.~\ref{fig:mod_D1_J1}(a) and (c), the $D$-modulation from the piezoelectric displacement provides nearly zero driving on the spin-qubit; however, we will retain it for consistency. Finally, we note that detuning of the resonance frequency $\Delta f = |\omega_0 - \omega_{ij}|$ increases the effective Rabi rate $\Omega' = \sqrt{ \Omega^2 + \Delta f^2 }$, although the amplitude of spin components is reduced to $\Omega/\Omega'$ as expected, as shown in~\cite{SM4}.

\begin{figure}
\centering
\includegraphics[width=1\linewidth]{  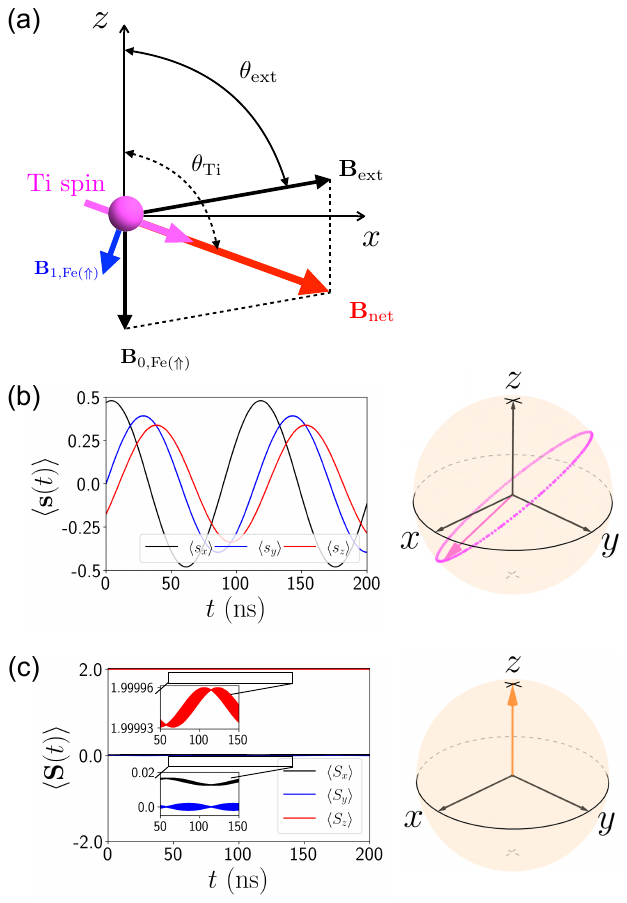}
    \caption{Spin dynamics of a Fe-Ti pair using $DJ$-modulation: (a) Illustration of the static net magnetic field composed of the external field $\bm{B_\textrm{ext}}$ and the field induced by the spin-up state of Fe, $\bm{B}_{0, \textrm{Fe} (\Uparrow)}$, acting on the Ti spin. $\bm{B}_{1, \textrm{Fe} (\Uparrow)}$ is the time-dependent driving field induced by Fe spin dynamics. Time-dependence of (b) the Ti spin in the rotating frame and (c) the Fe spin and their representations on a Bloch sphere. The arrows indicate the initial spin configurations of Ti and Fe. As we can see, the Fe spin does not undergo any measurable time dependence. The transition frequency is $\omega_{ij} /(2 \pi) \approx 26 \text{ GHz}$, and the Rabi rate is $\Omega / (2 \pi)= 9.41 \text{ MHz}$. Simulation parameters: $D_0 = -4.7$ meV, $D_1 = 0.41  \mu\text{eV}$, $J_0 = 6.8 \text{ GHz}$, $J_1=0.01 \text{ GHz}$. $B_\text{ext} = 0.9 \text{ T}$, $\theta_\text{ext}= 80^{\circ}$. We employ isotropic $g$-tensor $g=2$.}
    \label{fig:time_evolution}
\end{figure}

{\emph{Angle-dependence of the Rabi rate}--} Figure~\ref{fig:time_evolution} presents the Rabi driving process of a Fe-Ti pair. We first illustrate the net magnetic field on the position of the Ti spin, $\bm{B}_\text{net} = \bm{B}_\text{ext} + \bm{B}_{0, \text{Fe} (\Uparrow)}$ where $\bm{B}_{0, \text{Fe} (\Uparrow)}$ is the static magnetic field induced effectively by the spin-up state of Fe. The static fields are drawn proportionally with their real magnitude so that the initial direction of the Ti spin, which is parallel to $\bm{B}_\text{net}$, is self-explanatory. Thus, $\bm{B}_\text{net}$ constitutes the quantization axis where the Hamiltonian of the Ti alone is diagonal. Figures~\ref{fig:time_evolution}(b) and (c) show the time evolution of Ti and Fe spins. In Fig.~\ref{fig:time_evolution}(b) we plot the corresponding Rabi rotations in the Bloch sphere under resonant condition while Fig.~\ref{fig:time_evolution}(c) shows the evolution of the Fe under the same condition, remaining almost unchanged in its initial spin-up state. Nonetheless, the modulation also induces small oscillations in the Fe spin, as depicted in the inset of Fig.~\ref{fig:time_evolution}(c). Using realistic parameters (see figure caption), we obtain Rabi rate of Ti spin $\Omega/(2 \pi) \approx 10 \textrm{ MHz}$ {\corr{which compares well to experimental values of 10$-$30 MHz, depending on the specific driving condition}}~\cite{Phark2023Electric-Field-DrivenMagnet, Phark2023Double-ResonanceSurface,Bui2024All-electricalSpinb}.
\begin{figure}
\includegraphics[width=1\linewidth]{  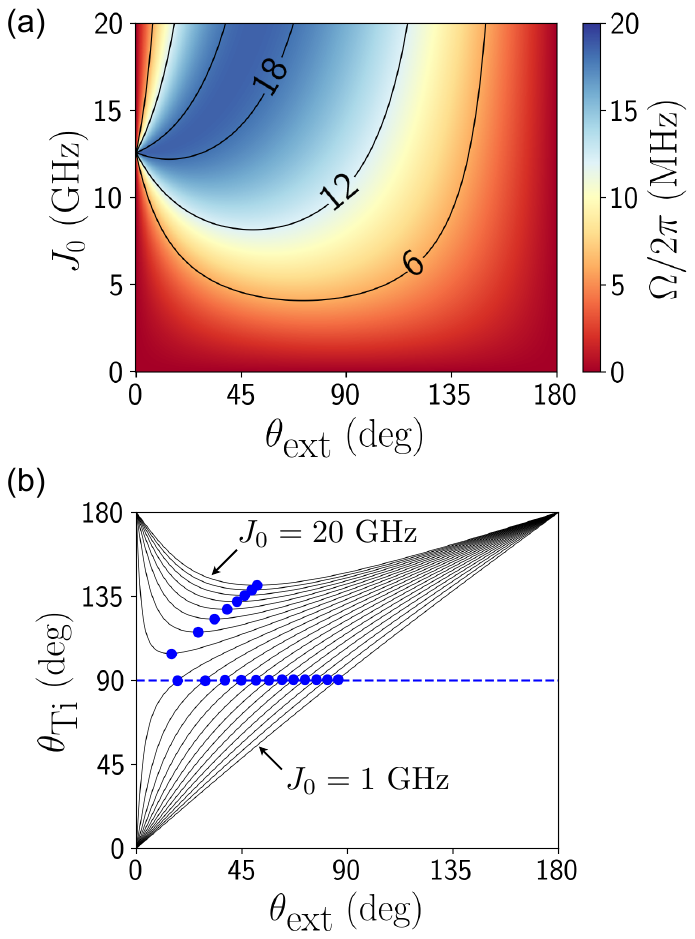}
\caption{Optimization of Rabi rates. (a) Dependence of Rabi rates on the static coupling strength $J_0$ and the external magnetic field angle $\theta_\text{ext}$. Here we use a fixed ratio of $J_1/J_0=1/680$.
(b) Dependence of the Ti spin's angle $\theta_\textrm{Ti}$ on the external magnetic field angle $\theta_\text{ext}$. Each line corresponds to a fixed value of $J_0$, ranging from 1 GHz to 20 GHz in steps of 1 GHz. Blue dots indicate the maximum Rabi rates. 
The remaining parameters are $B_\mathrm{ext}=0.9$ T, isotropic $g$-tensor $g=2$, $D_0=-4.7$ meV and $D_1=0.41\ \mu$eV.}
\label{fig:theta_dep}
\end{figure}
\vspace{.1cm}

Figure~\ref{fig:time_evolution}(a) also illustrates the induced driving field $\bm{B}_{1, \text{Fe}(\Uparrow)}$ from the spin-up state of the Fe. Its amplitude reads
\begin{equation}
\begin{aligned}
B_{1, \text{Fe}(\Uparrow)} &= B_{0, \text{Fe}(\Uparrow)} \frac{J_1}{J_0} \cos(\theta_\mathrm{Ti}-\pi/2)  \\ 
&=  \frac{\langle S_{z} \rangle J_1}{g\mu_{\rm B}} \sin(\theta_\mathrm{Ti}) , 
\end{aligned}
\label{eq:B1_Fe}
\end{equation}
where the polar angle of the initial Ti spin, $\theta_\textrm{Ti}$, can be determined via the static magnetic fields as follows

\begin{eqnarray*}\theta_\mathrm{Ti}=\cos^{-1}\left[ \frac{B_{\mathrm{ext}}\cos(\theta_{\mathrm{ext}})- B_{0, \text{Fe}(\Uparrow)}}{B_\mathrm{net}}\right],
\end{eqnarray*}
{\corr{where the minus sign in front of $ B_{0, \text{Fe}(\Uparrow)}$  originates from the negative sign in the exchange coupling term in Eq.~(\ref{eq:stat_hamil}).}}
{\corr{The strength of the driving field that determines the Rabi rate is the perpendicular projection}} of the driving field onto the Ti spin direction{\corr{, as indicated by its proportionality to $\sin(\theta_\mathrm{Ti})$. In this context,  Eq.}} (\ref{eq:B1_Fe}) highlights {\corr{a}} crucial factor for enhancing the Rabi rate: its behavior with respect to the angle $\theta_\mathrm{Ti}$. This angle depends on the external magnetic field; thus, for a given exchange coupling $J_0$, there is an optimal direction of the external magnetic field that enables the Fe to drive the Ti {\cor{most efficiently}}. {\corr{Notably, $\theta_{\mathrm{ext}}=0$ degrees results in $\theta_{\mathrm{Ti}}=0$ degrees and therefore, no driving field acts on the spin-qubit, emphasizing the importance of selecting the proper external magnetic field angle.}}

To {\corr{better}} illustrate {\corr{the role of $\theta_{\mathrm{ext}}$}}, Fig.~\ref{fig:theta_dep}(a) presents a color map of the Rabi rate as a function of the exchange coupling $J_0$ and the external magnetic field angle $\theta_{\mathrm{ext}}$. {\cor{Since $J_1\propto J_0$ we use a fixed ratio for the modulation strength of $J_1=J_0/680$, so}} that $J_0=6.8$ GHz results in a modulation of 10 MHz. The color map confirms Eq.~(\ref{eq:B1_Fe}): for a given static exchange coupling, the optimal magnetic field angle is the one that aligns the Ti at $90$ degrees, as indicated in Fig.~\ref{fig:theta_dep}(b), unless the magnetic field from the Fe is strongly dominating. In such scenarios, it is evident by Fig.~\ref{fig:theta_dep}(b) that the Rabi rate reaches its maximum value for an external field angle which aligns the Ti spin as close as possible to 90 degrees.
Since both $J_0$ and $\theta_\mathrm{ext}$ are adjustable parameters in the experiment, these findings open the door for optimizing the driving from a single-atom magnet. {\cor{In practice, this has already been implemented experimentally, as it is often necessary to tune $\theta_\mathrm{ext}$ to properly drive the spin qubit adjacent to the quantum magnet.}}
An {\cor{additional}} observation regarding Eq.~(\ref{eq:B1_Fe}) is the dependence on $\langle S_{z} \rangle$, suggesting that magnets with higher spin exhibit enhanced Rabi rates for a given modulation $J_1$. 

In summary, we have explored that in an atomic pair formed by a single-atom magnet and a spin-$1/2$ on a surface, ESR can be achieved by modulation of the crystal field anisotropy of the single-atom magnet or by modulation of the Fe-Ti exchange coupling. Modulation of the anisotropy alone results in Rabi rates that are 1$-$2 orders of magnitude smaller than those observed in the experiment~\cite{Phark2023Electric-Field-DrivenMagnet,Wang2023AnPlatform,Wang2023UniversalSurface}, however, even a small modulation of the exchange coupling results in a highly efficient Rabi process, with Rabi rates easily reaching tens of MHz in agreement with the experiment. 
We have also highlighted the significance of adjustable parameters, such as the static exchange coupling, the external magnetic field angle and the spin of the magnet in maximizing the Rabi rate.
We therefore conclude that surface spin qubits built by combining a single-atom magnet and a spin-1/2 can be {\cor{efficiently}} controlled purely by electrical fields that modulate the exchange coupling between the two spins.


\textit{Acknowledgments.}\textemdash This work was supported by the Institute for Basic Science (IBS-R027-D1). The authors would like to thank Fabio Donati and Michael E. Flatt\'e for stimulating discussions. 
NL thanks projects  PID2021-127917NB-I00 by MCIN/AEI/10.13039/501100011033, QUAN-000021-01 by Gipuzkoa Provincial Council, IT-1527-22 by Basque Government, 202260I187 by CSIC, ESiM project 101046364 by EU. J.R-G acknowledges support from the German Research Foundation [Deutsche Forschungsgemeinschaft (DFG)] under Projects No. 450396347. Views and opinions expressed are however those of the author(s) only and do not necessarily reflect those of the EU. Neither the EU nor the granting authority can be held responsible for them. 

\bibliography{references}

@misc{SM1,
  note = {See Supplemental Material, Sec.~\ref{sect:modulation_superexchange}, for details on the substrate-mediated superexchange mechanism and its modulation by electric fields. It includes Refs.~\cite{Takasan2019,Koch:1018555,Takasan2019,bae2018tipairs}.}
}

@misc{SM2,
  note = {See Supplemental Material, Sec.~\ref{sect:comparison}, for a comparison of $J$-modulation in several systems, displayed in a table for direct comparison. It includes Refs.~\cite{Tsymbal2012Spintronics:Magnets, Mankovsky_prb_2021, Juarez-Reyes2014Electric-field-modulated1, Matsukura2015ControlFieldsb, Dhingra_2022}.}
}

@misc{SM3,
  note = {See Supplemental Material, Sec.~\ref{sect:spatial_component}, for an analysis of the contribution of each component of the exchange coupling to the Rabi rates.}
}

@misc{SM4,
  note = {See Supplemental Material, Sec.~\ref{sect:detuning_phase}, where we demonstrate full quantum control through exchange interaction modulation by detuning and phase control.}
}

@misc{SM5,
  note = {See Supplemental Material, Sec.~\ref{sect:DFT_calculation}, where we present DFT results showing the dependence of displacement and exchange coupling on the electric field. It includes Refs.~\cite{Giannozzi_2009, Giannozzi_2017, dalcorso2014, C7NR01596H, Phark2023Electric-Field-DrivenMagnet, Phark2023Double-ResonanceSurface, Bui2024All-electricalSpinb, Seifert2020LongitudinalMicroscope, choi2017, Choi2017a, Schurkus2020, noodleman1981valence, Phark2023Electric-Field-DrivenMagnet, Sheng2020}.}
}

@article{bae2018tipairs,
author = {Y. Bae  and K. Yang  and P. Willke  and T. Choi  and A. J. Heinrich  and C. P. Lutz },
title = {Enhanced quantum coherence in exchange coupled spins via singlet-triplet transitions},
journal = {Science Advances},
volume = {4},
number = {11},
pages = {eaau4159},
year = {2018},
doi = {10.1126/sciadv.aau4159},
URL = {https://www.science.org/doi/abs/10.1126/sciadv.aau4159},
eprint = {https://www.science.org/doi/pdf/10.1126/sciadv.aau4159}}

@article{Giannozzi_2017,
doi = {10.1088/1361-648X/aa8f79},
url = {https://dx.doi.org/10.1088/1361-648X/aa8f79},
year = {2017},
month = {oct},
publisher = {IOP Publishing},
volume = {29},
number = {46},
pages = {465901},
author = {P Giannozzi and O Andreussi and T Brumme and O Bunau and M Buongiorno Nardelli and M Calandra and R Car and C Cavazzoni and D Ceresoli and M Cococcioni and N Colonna and I Carnimeo and A Dal Corso and S de Gironcoli and P Delugas and R A DiStasio and A Ferretti and A Floris and G Fratesi and G Fugallo and R Gebauer and U Gerstmann and F Giustino and T Gorni and J Jia and M Kawamura and H-Y Ko and A Kokalj and E Küçükbenli and M Lazzeri and M Marsili and N Marzari and F Mauri and N L Nguyen and H-V Nguyen and A Otero-de-la-Roza and L Paulatto and S Poncé and D Rocca and R Sabatini and B Santra and M Schlipf and A P Seitsonen and A Smogunov and I Timrov and T Thonhauser and P Umari and N Vast and X Wu and S Baroni},
title = {Advanced capabilities for materials modelling with Quantum ESPRESSO},
journal = {Journal of Physics: Condensed Matter}
}

@article{Giannozzi_2009,
doi = {10.1088/0953-8984/21/39/395502},
url = {https://dx.doi.org/10.1088/0953-8984/21/39/395502},
year = {2009},
month = {sep},
publisher = {},
volume = {21},
number = {39},
pages = {395502},
author = {Paolo Giannozzi and Stefano Baroni and Nicola Bonini and Matteo Calandra and Roberto Car and Carlo Cavazzoni and Davide Ceresoli and Guido L Chiarotti and Matteo Cococcioni and Ismaila Dabo and Andrea Dal Corso and Stefano de Gironcoli and Stefano Fabris and Guido Fratesi and Ralph Gebauer and Uwe Gerstmann and Christos Gougoussis and Anton Kokalj and Michele Lazzeri and Layla Martin-Samos and Nicola Marzari and Francesco Mauri and Riccardo Mazzarello and Stefano Paolini and Alfredo Pasquarello and Lorenzo Paulatto and Carlo Sbraccia and Sandro Scandolo and Gabriele Sclauzero and Ari P Seitsonen and Alexander Smogunov and Paolo Umari and Renata M Wentzcovitch},
title = {QUANTUM ESPRESSO: a modular and open-source software project for quantum
simulations of materials},
journal = {Journal of Physics: Condensed Matter}
}

@article{dalcorso2014,
title = {Pseudopotentials periodic table: From H to Pu},
journal = {Computational Materials Science},
volume = {95},
pages = {337-350},
year = {2014},
issn = {0927-0256},
doi = {https://doi.org/10.1016/j.commatsci.2014.07.043},
url = {https://www.sciencedirect.com/science/article/pii/S0927025614005187},
author = {Andrea {Dal Corso}}
}

@article{Schurkus2020,
    author = {Schurkus, Henry and Chen, Dian-Teng and Cheng, Hai-Ping and Chan, Garnet and Stanton, John},
    title = {Theoretical prediction of magnetic exchange coupling constants from broken-symmetry coupled cluster calculations},
    journal = {The Journal of Chemical Physics},
    volume = {152},
    number = {23},
    pages = {234115},
    year = {2020},
    month = {06},
    issn = {0021-9606},
    doi = {10.1063/1.5144696},
    url = {https://doi.org/10.1063/1.5144696},
}

@article{noodleman1981valence,
  title={Valence bond description of antiferromagnetic coupling in transition metal dimers},
  author={Noodleman, Louis},
  journal={The Journal of Chemical Physics},
  volume={74},
  number={10},
  pages={5737--5743},
  year={1981},
  publisher={American Institute of Physics}
}

@article{Sheng2020,
author = {Sheng, Xianghai and Thompson, Lee M. and Hratchian, Hrant P.},
title = {Assessing the Calculation of Exchange Coupling Constants and Spin Crossover Gaps Using the Approximate Projection Model To Improve Density Functional Calculations},
journal = {Journal of Chemical Theory and Computation},
volume = {16},
number = {1},
pages = {154-163},
year = {2020},
doi = {10.1021/acs.jctc.9b00387},
note ={PMID: 31743016},
URL = {https://doi.org/10.1021/acs.jctc.9b00387
}
}

@article{Wang2023_universal,
  author    = {Wang, Yu and Haze, Masahiro and Bui, Hong T. and Soe, We-hyo and Aubin, Herve and Ardavan, Arzhang and Heinrich, Andreas J. and Phark, Soo-hyon},
  title     = {Universal quantum control of an atomic spin qubit on a surface},
  journal   = {npj Quantum Information},
  year      = {2023},
  volume    = {9},
  number    = {1},
  pages     = {48},
  doi       = {10.1038/s41534-023-00716-6},
  url       = {https://doi.org/10.1038/s41534-023-00716-6},
  issn      = {2056-6387}
}

@Article{eric_unraveling_2025,
author ="Switzer, Eric and Reina Gálvez, Jose and Giedke, Geza and Rahman, Talat Shahnaz and Wolf, Christoph and Choi, Deungjang and Lorente, Nicolas",
title  ="Unraveling spin entanglement using quantum gates with scanning tunneling microscopy-driven electron spin resonance",
journal  ="Nanoscale Adv.",
year  ="2025",
pages  ="-",
publisher  ="RSC",
doi  ="10.1039/D5NA00421G",
url  ="http://dx.doi.org/10.1039/D5NA00421G"}

@article{kot2023electric,
  author    = {Kot, Piotr and Ismail, Maneesha and Drost, Robert and Siebrecht, Janis and Huang, Haonan and Ast, Christian R.},
  title     = {Electric control of spin transitions at the atomic scale},
  journal   = {Nature Communications},
  volume    = {14},
  pages     = {6612},
  year      = {2023},
  publisher = {Springer Nature},
  doi       = {10.1038/s41467-023-42287-2},
  url       = {https://doi.org/10.1038/s41467-023-42287-2}
}

@article{Huang2025,
  author    = {Wantong Huang and M{\'a}t{\'e} Stark and Paul Greule and Kwan Ho Au-Yeung and Daria Sostina and Jos{\'e} Reina G{\'a}lvez and Christoph S{\"u}rgers and Wolfgang Wernsdorfer and Christoph Wolf and Philip Willke},
  title     = {Quantum spin-engineering in on-surface molecular ferrimagnets},
  journal   = {Nature Communications},
  year      = {2025},
  volume    = {16},
  number    = {1},
  pages     = {5208},
  doi       = {10.1038/s41467-025-60409-w},
  url       = {https://doi.org/10.1038/s41467-025-60409-w},
  issn      = {2041-1723},
}

@misc{czap2025magneticresonanceimagingsingle,
      title={Magnetic Resonance Imaging of Single Organic Radicals with Sub-Molecular Resolution}, 
      author={Gregory Czap and Christoph Wolf and Jose Reina-Gálvez and Mark H. Sherwood and Christopher P. Lutz},
      year={2025},
      eprint={2504.18043},
      archivePrefix={arXiv},
      primaryClass={cond-mat.mes-hall},
      url={https://arxiv.org/abs/2504.18043}, 
}

@misc{greule2025spinelectriccontrolindividualmolecules,
      title={Spin-Electric Control of Individual Molecules on Surfaces}, 
      author={Paul Greule and Wantong Huang and Máté Stark and Kwan Ho Au-Yeung and Johannes Schwenk and Jose Reina-Gálvez and Christoph Sürgers and Wolfgang Wernsdorfer and Christoph Wolf and Philip Willke},
      year={2025},
      eprint={2507.13699},
      archivePrefix={arXiv},
      primaryClass={cond-mat.mes-hall},
      url={https://arxiv.org/abs/2507.13699}, 
}

@article{zhang_influence_2023,
author = {Zhang, Xue and Reina-Gálvez, Jose and Wolf, Christoph and Wang, Yu and Aubin, Herv{\'e} and Heinrich, Andreas J. and Choi, Taeyoung},
title = {Influence of the Magnetic Tip on Heterodimers in Electron Spin Resonance Combined with Scanning Tunneling Microscopy},
journal = {ACS Nano},
volume = {17},
number = {17},
pages = {16935-16942},
year = {2023},
doi = {10.1021/acsnano.3c04024},
    note ={PMID: 37643247},
URL = { https://doi.org/10.1021/acsnano.3c04024
},
eprint = { https://doi.org/10.1021/acsnano.3c04024}
}

@article{Furuya2021,
  title = {Control of superexchange interactions with DC electric fields},
  author = {Furuya, Shunsuke C. and Takasan, Kazuaki and Sato, Masahiro},
  journal = {Phys. Rev. Res.},
  volume = {3},
  issue = {3},
  pages = {033066},
  numpages = {24},
  year = {2021},
  month = {Jul},
  publisher = {American Physical Society},
  doi = {10.1103/PhysRevResearch.3.033066},
  url = {https://link.aps.org/doi/10.1103/PhysRevResearch.3.033066}
}

@INPROCEEDINGS{Koch:1018555,
      author       = {Koch, Erik},
      title        = {{E}xchange {M}echanisms},
      volume       = {13},
      address      = {Jülich},
      publisher    = {Forschungszentrum Jülich},
      reportid     = {FZJ-2023-04883},
      isbn         = {9783958066892},
      series       = {Modeling and Simulation},
      pages        = {5.1-31},
      year         = {2023},
      comment      = {Orbital Physics in Correlated Matter},
      booktitle     = {Orbital Physics in Correlated Matter},
      month         = {Sep},
      date          = {2023-09-18},
      organization  = {Autumn School on Correlated Electrons,
                       Jülich (Germany), 18 Sep 2023 - 22 Sep
                       2023},
      cin          = {JSC},
      cid          = {I:(DE-Juel1)JSC-20090406},
      pnm          = {5111 - Domain-Specific Simulation $\&$ Data Life Cycle Labs
                      (SDLs) and Research Groups (POF4-511)},
      pid          = {G:(DE-HGF)POF4-5111},
      typ          = {PUB:(DE-HGF)8 / PUB:(DE-HGF)7},
      doi          = {10.34734/FZJ-2023-04883},
      url          = {https://juser.fz-juelich.de/record/1018555},
}

@article{Donati2016atommagnet,
author = {F. Donati  and S. Rusponi  and S. Stepanow  and C. Wäckerlin  and A. Singha  and L. Persichetti  and R. Baltic  and K. Diller  and F. Patthey  and E. Fernandes  and J. Dreiser  and Z. Sljivancanin  and K. Kummer  and C. Nistor  and P. Gambardella  and H. Brune},
title = {Magnetic remanence in single atoms},
journal = {Science},
volume = {352},
number = {6283},
pages = {318-321},
year = {2016},
doi = {10.1126/science.aad9898}}

@article{Strandberg2010,
  title = {Magnetic interactions of substitutional Mn pairs in GaAs},
  author = {Strandberg, T. O. and Canali, C. M. and MacDonald, A. H.},
  journal = {Phys. Rev. B},
  volume = {81},
  issue = {5},
  pages = {054401},
  numpages = {19},
  year = {2010},
  month = {Feb},
  publisher = {American Physical Society},
  doi = {10.1103/PhysRevB.81.054401},
  url = {https://link.aps.org/doi/10.1103/PhysRevB.81.054401}
}

@article{Rodriguez2025,
  title = {Nonresonant Electric Quantum Control of Individual On-Surface Spins},
  author = {Rodr\'{\i}guez, S. A. and G\'omez, S. S. and Fern\'andez-Rossier, J. and Ferr\'on, A.},
  journal = {Phys. Rev. Lett.},
  volume = {134},
  issue = {5},
  pages = {056703},
  numpages = {6},
  year = {2025},
  month = {Feb},
  publisher = {American Physical Society},
  doi = {10.1103/PhysRevLett.134.056703},
  url = {https://link.aps.org/doi/10.1103/PhysRevLett.134.056703}
}

@article{Takasan2019,
  title = {Control of magnetic and topological orders with a DC electric field},
  author = {Takasan, Kazuaki and Sato, Masahiro},
  journal = {Phys. Rev. B},
  volume = {100},
  issue = {6},
  pages = {060408},
  numpages = {6},
  year = {2019},
  month = {Aug},
  publisher = {American Physical Society},
  doi = {10.1103/PhysRevB.100.060408},
  url = {https://link.aps.org/doi/10.1103/PhysRevB.100.060408}
}

@Article{C7NR01596H,
author ="Qin, Guangzhao and Qin, Zhenzhen and Yue, Sheng-Ying and Yan, Qing-Bo and Hu, Ming",
title  ="External electric field driving the ultra-low thermal conductivity of silicene",
journal  ="Nanoscale",
year  ="2017",
volume  ="9",
issue  ="21",
pages  ="7227-7234",
publisher  ="The Royal Society of Chemistry",
doi  ="10.1039/C7NR01596H",
url  ="http://dx.doi.org/10.1039/C7NR01596H",
}

@misc{zhang2024electricfieldcontrolexchange,
      title={Electric field control of the exchange field of a single spin impurity on a surface}, 
      author={Xue Zhang and Jose Reina-Gálvez and Di'an Wu and Jan Martinek and Andreas J. Heinrich and Taeyoung Choi and Christoph Wolf},
      year={2024},
      eprint={2412.03866},
      archivePrefix={arXiv},
      primaryClass={cond-mat.mes-hall},
      url={https://arxiv.org/abs/2412.03866}, 
}

@article{Switzer_PhysRevA_2021,
  title = {Anisotropy-exchange resonance as a mechanism for entangled state switching},
  author = {Switzer, Eric D. and Zhang, Xiao-Guang and Rahman, Talat S.},
  journal = {Phys. Rev. A},
  volume = {104},
  issue = {5},
  pages = {052434},
  numpages = {6},
  year = {2021},
  month = {Nov},
  publisher = {American Physical Society},
  doi = {10.1103/PhysRevA.104.052434},
  url = {https://link.aps.org/doi/10.1103/PhysRevA.104.052434}
}

@article{Switzer_2022,
doi = {10.1088/2399-6528/ac7e1d},
url = {https://dx.doi.org/10.1088/2399-6528/ac7e1d},
year = {2022},
month = {jul},
publisher = {IOP Publishing},
volume = {6},
number = {7},
pages = {075007},
author = {E D Switzer and X-G Zhang and T S Rahman},
title = {Electronic control and switching of entangled spin state using anisotropy and exchange in the three-particle paradigm},
journal = {Journal of Physics Communications}
}

@article{Dhingra_2022,
doi = {10.1088/1361-648X/ac8c11},
url = {https://dx.doi.org/10.1088/1361-648X/ac8c11},
year = {2022},
month = {sep},
publisher = {IOP Publishing},
volume = {34},
number = {44},
pages = {441501},
author = {Archit Dhingra and Xuedong Hu and Mario F Borunda and Joseph F Johnson and Christian Binek and Jonathan Bird and Alpha T N’Diaye and Jean-Pascal Sutter and Emilie Delahaye and Eric D Switzer and Enrique del Barco and Talat S Rahman and Peter A Dowben},
title = {Molecular transistors as substitutes for quantum information applications},
journal = {Journal of Physics: Condensed Matter}
}

@article{J_Cuevas_C_Ast_ESR_theory_2024,
  title = {Theory of electron spin resonance in scanning tunneling microscopy},
  author = {Ast, Christian R. and Kot, Piotr and Ismail, Maneesha and de-la-Pe\~na, Sebasti\'an and Fern\'andez-Dom\'{\i}nguez, Antonio I. and Cuevas, Juan Carlos},
  journal = {Phys. Rev. Res.},
  volume = {6},
  issue = {2},
  pages = {023126},
  numpages = {18},
  year = {2024},
  month = {May},
  publisher = {American Physical Society},
  doi = {10.1103/PhysRevResearch.6.023126},
  url = {https://link.aps.org/doi/10.1103/PhysRevResearch.6.023126}
}

@article{choi2017,
title = {Studies of magnetic dipolar interaction between individual atoms using ESR-STM},
journal = {Current Applied Physics},
volume = {17},
number = {11},
pages = {1513-1517},
year = {2017},
issn = {1567-1739},
doi = {https://doi.org/10.1016/j.cap.2017.08.011},
url = {https://www.sciencedirect.com/science/article/pii/S1567173917302298},
author = {Taeyoung Choi and Christopher P. Lutz and Andreas J. Heinrich}
}

@article{DFT_footnote,
  author = {},
  title = {},
  journal = {},
  year = {},
  note = {\hspace{-0.165cm}We perfomed plane-wave basis DFT calculations of a Fe-Fe and Fe-Ti dimer on 2 ML of MgO/4 ML of silver and extracted the exchange coupling energy as function of the applied electric field}
}

@article{Spin_torque-driven_Kovarik_2024,
author = {Stepan Kovarik  and Richard Schlitz  and Aishwarya Vishwakarma  and Dominic Ruckert  and Pietro Gambardella  and Sebastian Stepanow },
title = {Spin torque–driven electron paramagnetic resonance of a single spin in a pentacene molecule},
journal = {Science},
volume = {384},
number = {6702},
pages = {1368-1373},
year = {2024},
doi = {10.1126/science.adh4753},
URL = {https://www.science.org/doi/abs/10.1126/science.adh4753}}

@article{Tsymbal2012Spintronics:Magnets,
    title = {{Spintronics: Electric toggling of magnets}},
    year = {2012},
    journal = {Nature Materials},
    author = {Tsymbal, Evgeny Y.},
    number = {1},
    pages = {12--13},
    volume = {11},
    publisher = {Nature Publishing Group},
    url = {http://dx.doi.org/10.1038/nmat3205},
    doi = {10.1038/nmat3205},
    issn = {14764660}
}

@article{Choi2017a,
    title = {{Atomic-scale sensing of the magnetic dipolar field from single atoms}},
    year = {2017},
    journal = {Nature Nanotechnology},
    author = {Choi, Taeyoung and Paul, William and Rolf-Pissarczyk, Steffen and MacDonald, Andrew J. and Natterer, Fabian D. and Yang, Kai and Willke, Philip and Lutz, Christopher P. and Heinrich, Andreas J.},
    number = {5},
    pages = {420--424},
    volume = {12},
    publisher = {Nature Publishing Group},
    url = {http://dx.doi.org/10.1038/nnano.2017.18},
    isbn = {0031-9007},
    doi = {10.1038/nnano.2017.18},
    issn = {17483395},
    pmid = {20929842},
    arxivId = {1706.09793}
}

@article{Mankovsky_prb_2021,
    title = {{Electric-field control of exchange interactions}},
    year = {2021},
    journal = {Physical Review B},
    author = {Mankovsky, S. and Simon, E. and Polesya, S. and Marmodoro, A. and Ebert, H.},
    number = {17},
    month = {11},
    pages = {174443},
    volume = {104},
    publisher = {American Physical Society},
    url = {https://link.aps.org/doi/10.1103/PhysRevB.104.174443},
    doi = {10.1103/PhysRevB.104.174443},
    issn = {24699969},
    arxivId = {2108.00932}
}

@article{Matsukura2015ControlFieldsb,
    title = {{Control of magnetism by electric fields}},
    year = {2015},
    journal = {Nature Nanotechnology},
    author = {Matsukura, Fumihiro and Tokura, Yoshinori and Ohno, Hideo},
    number = {3},
    pages = {209--220},
    volume = {10},
    publisher = {Nature Publishing Group},
    doi = {10.1038/nnano.2015.22},
    issn = {17483395}
}

@article{Juarez-Reyes2014Electric-field-modulated1,
    title = {{Electric-field-modulated exchange coupling within and between magnetic clusters on metal surfaces: Mn dimers on Cu(1 1 1)}},
    year = {2014},
    journal = {Journal of Physics Condensed Matter},
    author = {Juarez-Reyes, L. and Stepanyuk, V. S. and Pastor, G. M.},
    number = {17},
    volume = {26},
    doi = {10.1088/0953-8984/26/17/176003},
    issn = {09538984},
    pmid = {24721806}
}

@article{Seifert2020LongitudinalMicroscope,
    title = {{Longitudinal and transverse electron paramagnetic resonance in a scanning tunneling microscope}},
    year = {2020},
    journal = {Science Advances},
    author = {Seifert, Tom S. and Kovarik, Stepan and Juraschek, Dominik M. and Spaldin, Nicola A. and Gambardella, Pietro and Stepanow, Sebastian},
    number = {40},
    month = {10},
    pages = {1--12},
    volume = {6},
    url = {http://advances.sciencemag.org/content/6/40/eabc5511.abstract},
    doi = {10.1126/sciadv.abc5511},
    issn = {23752548},
    pmid = {32998882},
    arxivId = {2005.07455}
}

@article{Troiani2019ManipulationAnisotropy,
    title = {{Manipulation of spin cluster qubits by electric field induced modulation of exchange coupling, g-factor, and axial anisotropy}},
    year = {2019},
    journal = {Physical Review B},
    author = {Troiani, Filippo},
    number = {15},
    pages = {155424},
    volume = {100},
    publisher = {American Physical Society},
    url = {https://doi.org/10.1103/PhysRevB.100.155424},
    doi = {10.1103/PhysRevB.100.155424},
    issn = {24699969},
    arxivId = {1907.08391}
}

@article{Wang2023UniversalSurface,
    title = {{Universal quantum control of an atomic spin qubit on a surface}},
    year = {2023},
    journal = {npj Quantum Information},
    author = {Wang, Yu and Haze, Masahiro and Bui, Hong T. and Soe, We hyo and Aubin, Herve and Ardavan, Arzhang and Heinrich, Andreas J. and Phark, Soo hyon},
    number = {1},
    pages = {3--8},
    volume = {9},
    publisher = {Springer US},
    doi = {10.1038/s41534-023-00716-6},
    issn = {20566387}
}

@article{reale2023erbium,
    title = {{Erbium and thulium on MgO(100)/Ag(100) as candidates for single atom qubits}},
    year = {2023},
    journal = {Physical Review B},
    author = {Reale, S. and Singha, A. and Ahmed, S. L. and Krylov, D. and Colazzo, L. and Wolf, C. and Casari, C. S. and Barla, A. and Fernandes, E. and Patthey, F. and Pivetta, M. and Rusponi, S. and Brune, H. and Donati, F.},
    number = {4},
    pages = {45427},
    volume = {107},
    publisher = {American Physical Society},
    doi = {10.1103/PhysRevB.107.045427},
    issn = {24699969}
}

@article{Phark2023Double-ResonanceSurface,
    title = {{Double-Resonance Spectroscopy of Coupled Electron Spins on a Surface}},
    year = {2023},
    journal = {ACS Nano},
    author = {Phark, Soo Hyon and Chen, Yi and Bui, Hong T. and Wang, Yu and Haze, Masahiro and Kim, Jinkyung and Bae, Yujeong and Heinrich, Andreas J. and Wolf, Christoph},
    number = {14},
    pages = {14144--14151},
    volume = {17},
    doi = {10.1021/acsnano.3c04754},
    issn = {1936086X},
    pmid = {37406167}
}

@article{Wang2023AnPlatform,
    title = {{An atomic-scale multi-qubit platform}},
    year = {2023},
    journal = {Science},
    author = {Wang, Yu and Chen, Yi and Bui, Hong T. and Wolf, Christoph and Haze, Masahiro and Mier, Cristina and Kim, Jinkyung and Choi, Deung Jang and Lutz, Christopher P. and Bae, Yujeong and Phark, Soo Hyon and Heinrich, Andreas J.},
    number = {6666},
    pages = {87--92},
    volume = {382},
    doi = {10.1126/science.ade5050},
    issn = {10959203},
    pmid = {37797000}
}

@article{Ferron2019Anisotropy,
    title = {{ Single spin resonance driven by electric modulation of the  g  -factor anisotropy }},
    year = {2019},
    journal = {Physical Review Research},
    author = {Ferr{\'{o}}n, A. and Rodr{\'{i}}guez, S. A. and G{\'{o}}mez, S. S. and Lado, J. L. and Fern{\'{a}}ndez-Rossier, J.},
    number = {3},
    pages = {1--13},
    volume = {1},
    doi = {10.1103/physrevresearch.1.033185},
    arxivId = {1909.07942}
}

@article{GeorgePRL2012,
  title = {Coherent Spin Control by Electrical Manipulation of the Magnetic Anisotropy},
  author = {George, Richard E. and Edwards, James P. and Ardavan, Arzhang},
  journal = {Phys. Rev. Lett.},
  volume = {110},
  issue = {2},
  pages = {027601},
  numpages = {5},
  year = {2013},
  month = {Jan},
  publisher = {American Physical Society},
  doi = {10.1103/PhysRevLett.110.027601},
  url = {https://link.aps.org/doi/10.1103/PhysRevLett.110.027601}
}

@article{Oba2015PRL,
  title = {Electric-Field-Induced Modification of the Magnon Energy, Exchange Interaction, and Curie Temperature of Transition-Metal Thin Films},
  author = {Oba, M. and Nakamura, K. and Akiyama, T. and Ito, T. and Weinert, M. and Freeman, A. J.},
  journal = {Phys. Rev. Lett.},
  volume = {114},
  issue = {10},
  pages = {107202},
  numpages = {5},
  year = {2015},
  month = {Mar},
  publisher = {American Physical Society},
  doi = {10.1103/PhysRevLett.114.107202},
  url = {https://link.aps.org/doi/10.1103/PhysRevLett.114.107202}
}

@article{Ishibashi_2018,
doi = {10.7567/APEX.11.063002},
url = {https://dx.doi.org/10.7567/APEX.11.063002},
year = {2018},
month = {may},
publisher = {The Japan Society of Applied Physics},
volume = {11},
number = {6},
pages = {063002},
author = {Mio Ishibashi and Kihiro T. Yamada and Yoichi Shiota and Fuyuki Ando and Tomohiro Koyama and Haruka Kakizakai and Hayato Mizuno and Kazumoto Miwa and Shimpei Ono and Takahiro Moriyama and Daichi Chiba and Teruo Ono},
title = {Electric field effect on exchange interaction in ultrathin Co films with ionic liquids},
journal = {Applied Physics Express}
}

@article{Fittipaldi_NatMat2019_Electricfieldmodulation,
	author = {Fittipaldi, Maria and Cini, Alberto and Annino, Giuseppe and Vindigni, Alessandro and Caneschi, Andrea and Sessoli, Roberta},
	date = {2019/04/01},
	doi = {10.1038/s41563-019-0288-5},
	id = {Fittipaldi2019},
	isbn = {1476-4660},
	journal = {Nature Materials},
	number = {4},
	pages = {329--334},
	title = {Electric field modulation of magnetic exchange in molecular helices},
	url = {https://doi.org/10.1038/s41563-019-0288-5},
	volume = {18},
	year = {2019}}

@article{Baadji_NatMat2009_electrostatic,
	author = {Baadji, Nadjib and Piacenza, Manuel and Tugsuz, Tugba and Sala, Fabio Della and Maruccio, Giuseppe and Sanvito, Stefano},
	date = {2009/10/01},
	doi = {10.1038/nmat2525},
	id = {Baadji2009},
	isbn = {1476-4660},
	journal = {Nature Materials},
	number = {10},
	pages = {813--817},
	title = {Electrostatic spin crossover effect in polar magnetic molecules},
	url = {https://doi.org/10.1038/nmat2525},
	volume = {8},
	year = {2009}}

@article{Islam_PRB2009_Cu3magnetic,
	author = {Islam, M. Fhokrul and Nossa, Javier F. and Canali, Carlo M. and Pederson, Mark},
	doi = {10.1103/PhysRevB.82.155446},
	issue = {15},
	journal = {Phys. Rev. B},
	month = {Oct},
	numpages = {9},
	pages = {155446},
	publisher = {American Physical Society},
	title = {First-principles study of spin-electric coupling in a ${{\text{Cu}}_{3}}$ single molecular magnet},
	url = {https://link.aps.org/doi/10.1103/PhysRevB.82.155446},
	volume = {82},
	year = {2010}}

@article{Phark2023Electric-Field-DrivenMagnet,
    title = {{Electric-Field-Driven Spin Resonance by On-Surface Exchange Coupling to a Single-Atom Magnet}},
    year = {2023},
    journal = {Advanced Science},
    author = {Phark, Soo hyon and Bui, Hong Thi and Ferr{\'{o}}n, Alejandro and Fern{\'{a}}ndez-Rossier, Joaquin and Reina-G{\'{a}}lvez, Jose and Wolf, Christoph and Wang, Yu and Yang, Kai and Heinrich, Andreas J. and Lutz, Christopher P.},
    number = {27},
    pages = {1--8},
    volume = {10},
    doi = {10.1002/advs.202302033},
    issn = {21983844},
    pmid = {37466177}
}

@article{Bui2024All-electricalSpinb,
    title = {{All-electrical driving and probing of dressed states in a single spin}},
    year = {2024},
    journal = {ACS Nano},
    author = {Bui, Hong T. and Wolf, Christoph and Wang, Yu and Haze, Masahiro and Ardavan, Arzhang and Heinrich, Andreas J. and Phark, Soo-hyon},
    number = {19},
    month = {5},
    pages = {12187--12193},
    volume = {18},
    publisher = {American Chemical Society},
    url = {http://arxiv.org/abs/2401.15440},
    doi = {10.1021/acsnano.4c00196},
    issn = {1936086X},
    arxivId = {2401.15440}
}

@misc{Rabi_paper,
      title={Contrasting exchange-field and spin-transfer torque driving mechanisms in all-electric electron spin resonance}, 
      author={Jose Reina-Galvez and Matyas Nachtigall and Nicolas Lorente and Jan Martinek and Christoph Wolf},
      year={2025},
      eprint={2503.24046},
      archivePrefix={arXiv},
      primaryClass={cond-mat.mes-hall},
      url={https://arxiv.org/abs/2503.24046}, 
}

@article{J_Reina_Galvez_2023,
	author = {Reina-G\'alvez, Jose and Wolf, Christoph and Lorente, Nicol\'as},
	doi = {10.1103/PhysRevB.107.235404},
	issue = {23},
	journal = {Phys. Rev. B},
	month = {Jun},
	numpages = {8},
	pages = {235404},
	publisher = {American Physical Society},
	title = {Many-body nonequilibrium effects in all-electric electron spin resonance},
	url = {https://link.aps.org/doi/10.1103/PhysRevB.107.235404},
	volume = {107},
	year = {2023}}

@article{Jinkyung_Hyperfine_2022,
	author = {Kim, Jinkyung and Noh, Kyungju and Chen, Yi and Donati, Fabio and Heinrich, Andreas J. and Wolf, Christoph and Bae, Yujeong},
	doi = {10.1021/acs.nanolett.2c02782},
	eprint = {https://doi.org/10.1021/acs.nanolett.2c02782},
	journal = {Nano Letters},
	note = {PMID: 36317830},
	number = {23},
	pages = {9766-9772},
	title = {Anisotropic Hyperfine Interaction of Surface-Adsorbed Single Atoms},
	url = {https://doi.org/10.1021/acs.nanolett.2c02782},
	volume = {22},
	year = {2022}}

@article{kovarik_electron_2022,
	author = {Kovarik, Stepan and Robles, Roberto and Schlitz, Richard and Seifert, Tom Sebastian and Lorente, Nicolas and Gambardella, Pietro and Stepanow, Sebastian},
	doi = {10.1021/acs.nanolett.2c00980},
	issn = {1530-6984},
	journal = {Nano Letters},
	month = may,
	note = {Publisher: American Chemical Society},
	number = {10},
	pages = {4176--4181},
	title = {Electron {Paramagnetic} {Resonance} of {Alkali} {Metal} {Atoms} and {Dimers} on {Ultrathin} {MgO}},
	url = {https://doi.org/10.1021/acs.nanolett.2c00980},
	urldate = {2022-07-06},
	volume = {22},
	year = {2022}}

@article{zhang_electron_2022,
    author = {Zhang, X. and Wolf, C. and Wang, Y. and Aubin, H. and Bilgeri, T. and Willke, P. and Heinrich, A. J. and Choi, T.},
    title = {Electron spin resonance of single iron phthalocyanine molecules and role of their non-localized spins in magnetic interactions},
    journal = {Nat. Chem.},
    volume = {14},
    number = {1},
    pages = {59-65},
    year = {2021},
    doi = {10.1038/s41557-021-00827-7}
}

@article{J_Reina_Galvez_2021,
	author = {Reina-G\'alvez, Jose and Lorente, Nicol\'as and Delgado, Fernando and Arrachea, Liliana},
	doi = {10.1103/PhysRevB.104.245435},
	issue = {24},
	journal = {Phys. Rev. B},
	month = {Dec},
	numpages = {16},
	pages = {245435},
	publisher = {American Physical Society},
	title = {All-electric electron spin resonance studied by means of Floquet quantum master equations},
	url = {https://link.aps.org/doi/10.1103/PhysRevB.104.245435},
	volume = {104},
	year = {2021}}

@article{Farinacci2022,
author = {Farinacci, Laëtitia and Veldman, Lukas M. and Willke, Philip and Otte, Sander},
title = {Experimental Determination of a Single Atom Ground State Orbital through Hyperfine Anisotropy},
journal = {Nano Letters},
volume = {22},
number = {21},
pages = {8470-8474},
year = {2022},
doi = {10.1021/acs.nanolett.2c02783},
    note ={PMID: 36305860},

URL = { 
    
        https://doi.org/10.1021/acs.nanolett.2c02783
    
    

},
eprint = { 
    
        https://doi.org/10.1021/acs.nanolett.2c02783
    
    

}

}

@article{Steinbrecher_Weerdenburg_2020,
	author = {Steinbrecher, Manuel and van Weerdenburg, Werner M. J. and Walraven, Etienne F. and van Mullekom, Niels P. E. and Gerritsen, Jan W. and Natterer, Fabian D. and Badrtdinov, Danis I. and Rudenko, Alexander N. and Mazurenko, Vladimir V. and Katsnelson, Mikhail I. and et al.},
	doi = {10.1103/physrevb.103.155405},
	issn = {2469-9969},
	journal = {Physical Review B},
	month = {Apr},
	number = {15},
	publisher = {American Physical Society (APS)},
	title = {Quantifying the interplay between fine structure and geometry of an individual molecule on a surface},
	url = {http://dx.doi.org/10.1103/PhysRevB.103.155405},
	volume = {103},
	year = {2021}}

@article{Weerdenburg_Steinbrecher_2020,
	author = {van Weerdenburg,Werner M. J. and Steinbrecher,Manuel and van Mullekom,Niels P. E. and Gerritsen,Jan W. and von Allw{\"o}rden,Henning and Natterer,Fabian D. and Khajetoorians,Alexander A.},
	doi = {10.1063/5.0040011},
	journal = {Review of Scientific Instruments},
	number = {3},
	pages = {033906},
	title = {A scanning tunneling microscope capable of electron spin resonance and pump--probe spectroscopy at mK temperature and in vector magnetic field},
	url = {https://doi.org/10.1063/5.0040011},
	volume = {92},
	year = {2021}}

@article{Yang_Paul_prl_2019,
	author = {Yang, Kai and Paul, William and Natterer, Fabian D. and Lado, Jose L. and Bae, Yujeong and Willke, Philip and Choi, Taeyoung and Ferr\'on, Alejandro and Fern\'andez-Rossier, Joaqu\'{\i}n and Heinrich, Andreas J. and Lutz, Christopher P.},
	doi = {10.1103/PhysRevLett.122.227203},
	issue = {22},
	journal = {Phys. Rev. Lett.},
	month = {Jun},
	numpages = {6},
	pages = {227203},
	publisher = {American Physical Society},
	title = {Tuning the Exchange Bias on a Single Atom from 1 mT to 10 T},
	url = {https://link.aps.org/doi/10.1103/PhysRevLett.122.227203},
	volume = {122},
	year = {2019}}

@article{Willke_Yang_natphys_2019,
	author = {Willke, Philip and Yang, Kai and Bae, Yu and Heinrich, Andreas and Lutz, Christopher},
	doi = {10.1038/s41567-019-0573-x},
	journal = {Nature Physics},
	month = {10},
	pages = {1005-1010},
	title = {Magnetic resonance imaging of single atoms on a surface},
	volume = {15},
	year = {2019}}

@article{Willke_Singha_nanolett_2019,
	author = {Willke, Philip and Singha, Aparajita and Zhang, Xue and Esat, Taner and Lutz, Christopher P. and Heinrich, Andreas J. and Choi, Taeyoung},
	doi = {10.1021/acs.nanolett.9b03559},
	journal = {Nano Letters},
	note = {PMID: 31661282},
	number = {11},
	pages = {8201-8206},
	title = {Tuning Single-Atom Electron Spin Resonance in a Vector Magnetic Field},
	url = {https://doi.org/10.1021/acs.nanolett.9b03559},
	volume = {19},
	year = {2019}}

@article{yang_coherent_2019,
	author = {Yang, Kai and Paul, William and Phark, Soo-Hyon and Willke, Philip and Bae, Yujeong and Choi, Taeyoung and Esat, Taner and Ardavan, Arzhang and Heinrich, Andreas J. and Lutz, Christopher P.},
	doi = {10.1126/science.aay6779},
	issn = {0036-8075, 1095-9203},
	journal = {Science},
	month = oct,
	number = {6464},
	pages = {509--512},
	title = {Coherent spin manipulation of individual atoms on a surface},
	url = {http://www.sciencemag.org/lookup/doi/10.1126/science.aay6779},
	urldate = {2020-01-23},
	volume = {366},
	year = {2019}}

@article{Seifert_Kovarik_pr_2020,
	author = {Seifert, T. S. and Kovarik, S. and Nistor, C. and Persichetti, L. and Stepanow, S. and Gambardella, P.},
	doi = {10.1103/PhysRevResearch.2.013032},
	issue = {1},
	journal = {Phys. Rev. Research},
	month = {Jan},
	numpages = {12},
	pages = {013032},
	publisher = {American Physical Society},
	title = {Single-atom electron paramagnetic resonance in a scanning tunneling microscope driven by a radio-frequency antenna at 4 K},
	url = {https://link.aps.org/doi/10.1103/PhysRevResearch.2.013032},
	volume = {2},
	year = {2020}}

@article{Seifert_Kovarik_eabc_2020,
	author = {Seifert, Tom S. and Kovarik, Stepan and Juraschek, Dominik M. and Spaldin, Nicola A. and Gambardella, Pietro and Stepanow, Sebastian},
	doi = {10.1126/sciadv.abc5511},
	elocation-id = {eabc5511},
	journal = {Science Advances},
	number = {40},
	publisher = {American Association for the Advancement of Science},
	title = {Longitudinal and transverse electron paramagnetic resonance in a scanning tunneling microscope},
	url = {https://advances.sciencemag.org/content/6/40/eabc5511},
	volume = {6},
	year = {2020}}

@article{J_Reina_Galvez_2019,
	author = {Reina G\'alvez, J. and Wolf, C. and Delgado, F. and Lorente, N.},
	doi = {10.1103/PhysRevB.100.035411},
	issue = {3},
	journal = {Phys. Rev. B},
	month = {Jul},
	numpages = {10},
	pages = {035411},
	publisher = {American Physical Society},
	title = {Cotunneling mechanism for all-electrical electron spin resonance of single adsorbed atoms},
	url = {https://link.aps.org/doi/10.1103/PhysRevB.100.035411},
	volume = {100},
	year = {2019}}

@article{Wolf_C_and_Delgado_F_2020,
	author = {Wolf, Christoph and Delgado, Fernando and Reina, Jos\'e and Lorente, Nicol{\'a}s},
	doi = {10.1021/acs.jpca.9b10749},
	journal = {The Journal of Physical Chemistry A},
	note = {PMID: 32098473},
	number = {11},
	pages = {2318-2327},
	title = {Efficient Ab Initio Multiplet Calculations for Magnetic Adatoms on MgO},
	url = {https://doi.org/10.1021/acs.jpca.9b10749},
	volume = {124},
	year = {2020}}

@article{Willke_Bae_science_2018,
	author = {Willke, Philip and Bae, Yujeong and Yang, Kai and Lado, Jose L. and Ferr{\'o}n, Alejandro and Choi, Taeyoung and Ardavan, Arzhang and Fern{\'a}ndez-Rossier, Joaqu{\'\i}n and Heinrich, Andreas J. and Lutz, Christopher P.},
	doi = {10.1126/science.aat7047},
	issn = {0036-8075},
	journal = {Science},
	number = {6412},
	pages = {336--339},
	publisher = {American Association for the Advancement of Science},
	title = {Hyperfine interaction of individual atoms on a surface},
	url = {http://science.sciencemag.org/content/362/6412/336},
	volume = {362},
	year = {2018}}

@article{Willke_Paul_sciadv_2018,
	author = {Willke, Philip and Paul, William and Natterer, Fabian D. and Yang, Kai and Bae, Yujeong and Choi, Taeyoung and Fern{\'a}ndez-Rossier, Joaquin and Heinrich, Andreas J. and Lutz, Christoper P.},
	doi = {10.1126/sciadv.aaq1543},
	elocation-id = {eaaq1543},
	journal = {Science Advances},
	number = {2},
	publisher = {American Association for the Advancement of Science},
	title = {Probing quantum coherence in single-atom electron spin resonance},
	volume = {4},
	year = {2018}}

@article{Yang_Bae_prl_2017,
	author = {Yang, Kai and Bae, Yujeong and Paul, William and Natterer, Fabian D. and Willke, Philip and Lado, Jose L. and Ferr\'on, Alejandro and Choi, Taeyoung and Fern\'andez-Rossier, Joaqu\'{\i}n and Heinrich, Andreas J. and Lutz, Christopher P.},
	doi = {10.1103/PhysRevLett.119.227206},
	issue = {22},
	journal = {Phys. Rev. Lett.},
	month = {Nov},
	numpages = {5},
	pages = {227206},
	publisher = {American Physical Society},
	title = {Engineering the Eigenstates of Coupled Spin-$1/2$ Atoms on a Surface},
	url = {https://link.aps.org/doi/10.1103/PhysRevLett.119.227206},
	volume = {119},
	year = {2017}}

@article{Y_Bae_advanced_science_2018,
	author = {Y. Bae and K. Yang and P. Willke and T. Choi and A. J. Heinrich and C. P. Lutz},
	doi = {10.1126/sciadv.aau4159},
	journal = {Science Advances},
	number = {11},
	pages = {eaau4159},
	title = {Enhanced quantum coherence in exchange coupled spins via singlet-triplet transitions},
	url = {https://www.science.org/doi/abs/10.1126/sciadv.aau4159},
	volume = {4},
	year = {2018}}

@article{Lado_Ferron_prb_2017,
    title = {{Exchange mechanism for electron paramagnetic resonance of individual adatoms}},
    year = {2017},
    journal = {Physical Review B},
    author = {Lado, J. L. and Ferr{\'{o}}n, A. and Fern{\'{a}}ndez-Rossier, J.},
    number = {20},
    month = {11},
    pages = {205420},
    volume = {96},
    publisher = {American Physical Society},
    url = {https://link.aps.org/doi/10.1103/PhysRevB.96.205420},
    doi = {10.1103/PhysRevB.96.205420},
    issn = {24699969}
}

@article{Natterer_Yang_nature_2017,
	author = {Natterer, Fabian D and Yang, Kai and Paul, William and Willke, Philip and Choi, Taeyoung and Greber, Thomas and Heinrich, Andreas J and Lutz, Christopher P},
	journal = {Nature},
	number = {7644},
	pages = {226--228},
	publisher = {Nature Research},
	title = {Reading and writing single-atom magnets},
	volume = {543},
	year = {2017}}

@article{Braun_Konig_prb_2004,
    author = {Braun, M. and K{\"o}nig, J. and Martinek, J.},
    title = {Theory of transport through quantum-dot spin valves in the weak-coupling regime},
    journal = {Phys. Rev. B},
    volume = {70},
    number = {19},
    pages = {195345},
    year = {2004},
    doi = {10.1103/PhysRevB.70.195345}
}

@article{Paul_Yang_natphys_2017,
	author = {Paul, William and Yang, Kai and Baumann, Susanne and Romming, Niklas and Choi, Taeyoung and Lutz, Christopher P and Heinrich, Andreas J},
    doi = {https://doi.org/10.1038/nphys3965},
	journal = {Nature Physics},
	number = {4},
	pages = {403--407},
	publisher = {Nature Research},
	title = {Control of the millisecond spin lifetime of an electrically probed atom},
	volume = {13},
	year = {2017}}

@article{Baumann_Paul_science_2015,
	author = {Baumann, Susanne and Paul, William and Choi, Taeyoung and Lutz, Christopher P. and Ardavan, Arzhang and Heinrich, Andreas J.},
	doi = {10.1126/science.aac8703},
	journal = {Science},
	number = {6259},
	pages = {417-420},
	title = {Electron paramagnetic resonance of individual atoms on a surface},
	volume = {350},
	year = {2015}}

\newpage
\onecolumngrid
\renewcommand{\thesection}{S\arabic{section}}
\renewcommand{\thetable}{S\arabic{table}}
\renewcommand{\thefigure}{S\arabic{figure}}
\renewcommand{\theequation}{S\arabic{equation}}
\setcounter{figure}{0}
\setcounter{table}{0}
\setcounter{section}{0}
\setcounter{equation}{0}

\section{Modulation of exchange coupling based on superexchange picture}
\label{sect:modulation_superexchange}

In this section, we explain the mechanism of exchange coupling modulation under AC electric field using tight-binding model. Our derivations follows Refs.~\cite{Furuya2021,Koch:1018555}. The system consists of a pair of spins
placed on a thin insulating layer. From the orbital perspective, the system can be seen as two $d$-orbitals localized at the magnetic atoms  and $p$-orbital from insulating layer acts as the ligand mediating the superexchange coupling between the two spins. As pointed out in other works the mechanism proposed here can be applied to both, direct exchange~\cite{Takasan2019} and superexchange~\cite{Furuya2021}. 

\subsection{Three orbital superexchange model}
\begin{figure}    
	\centering\includegraphics[width=\linewidth]{ 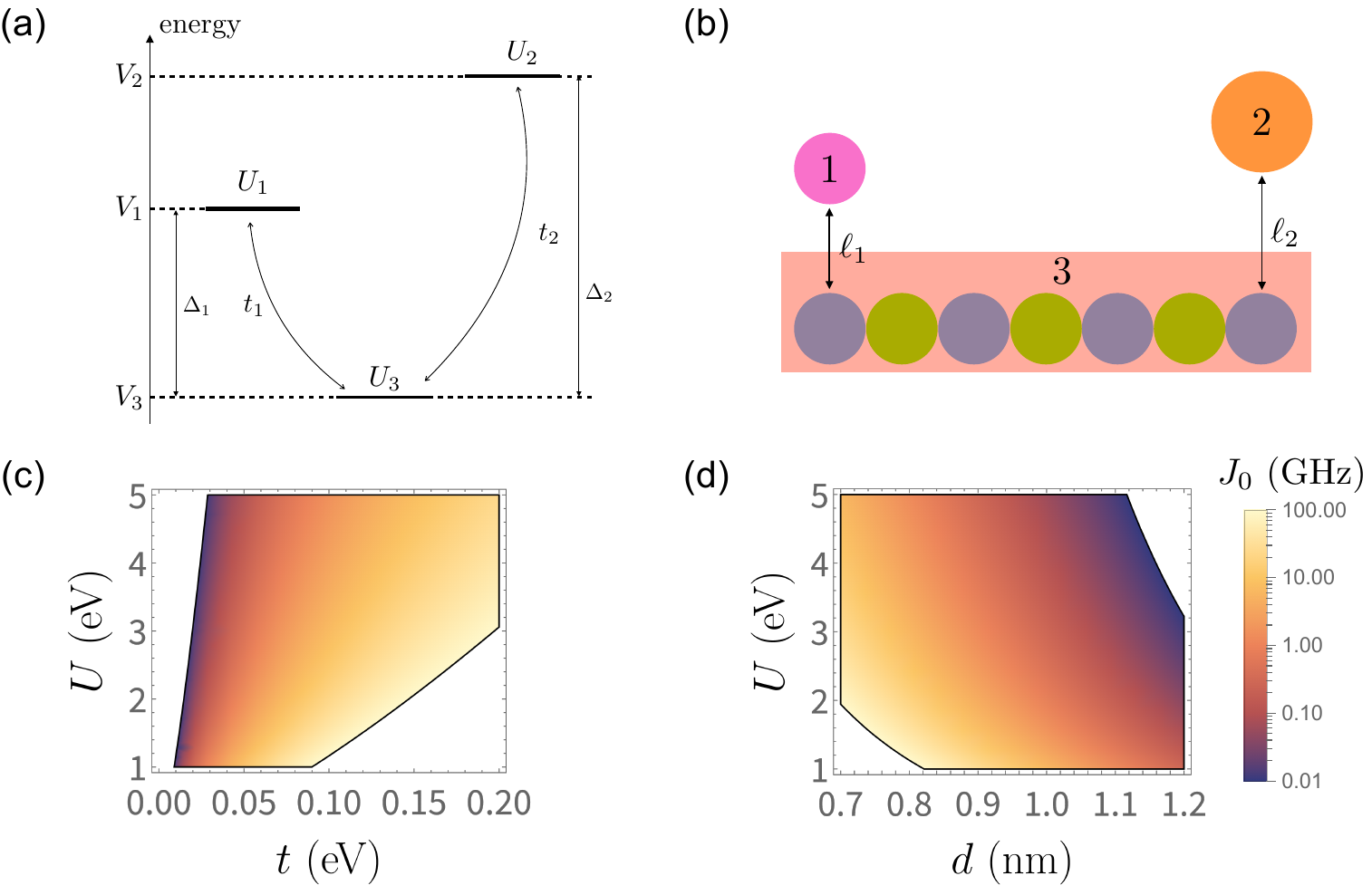}
	\caption{\textbf{Three-orbital superexchange model:} (a) Energy scheme of orbitals. Orbitals 1 and 2 describe the spin centers (i.e. adsorbates), whilst 3 is the effective substrate orbital. (b) Real-space schematic of the orbitals where we treat the insulating layer as an effective orbital $3$. $\ell_{1(2)}$ denotes the distance between orbitals and $\ell \equiv \ell_1 + \ell_2$. We calculated the absolute values of the exchange coupling $J_0$ from our model. (c) $J_0$ in log-scale as function of the on-site repulsion $U$ and hopping values $t = t_1 = t_2$. (d) The same simulation but $t$ was converted in atom-dimer distance $d$ assuming an exponential dependence of $t=t_0 \exp{(-d/4\lambda)}$ with $\lambda$ being the decay length. In all calculations $\Delta_1 = \Delta_2 = \Delta = 0.1 \text{ eV}$ and only the region with $ 0.01 \text{ GHz}\leq J_0 \leq 100 \text{ GHz}$ is shown.} \label{fig:tight_binding}
\end{figure}

We treat the connecting $p$ orbitals as a single effective orbital, denoted by number 3. 
The two spins are denoted as numbers 1 and 2. $c^\dagger_{k, \sigma}$ denotes for the creation operator of a spin-$\sigma$ in the orbital $i$ and $n_{k, \sigma} \equiv c^\dagger_{k, \sigma} c_{k, \sigma}$ is the corresponding number operator. 
The operators for other orbitals are defined likewise. The tight-binding Hamiltonian of the system can be written as
\begin{equation}
	H_\text{tb} = \sum_{k = 1}^3  U_k n_{k, \uparrow} n_{k, \downarrow} + \sum_{k=1}^3 \sum_{\sigma = \uparrow, \downarrow} V_k n_{k, \sigma} - \sum_{\sigma= \uparrow, \downarrow} \left( t_1 c^\dagger_{1, \sigma} c_{3,\sigma} + t_2 c^\dagger_{2,\sigma} c_{3,\sigma} + \text{h.c.} \right),
\end{equation}
where $U_k$ are on-site Coulomb repulsions and $V_k$ site energies of electron occupied orbital $k$. $t_{1(2)}$ is the hopping amplitude from orbital 1(2) to 3 (see Fig.~\ref{fig:tight_binding}(a)). 
Among these parameters, the site energies are affected by external electric field $\mathbf{E}$. We define the following energy difference of orbitals $1,2$ from the orbital $3$ excluding Coulomb repulsion as
\begin{equation}
	\Delta_k (\mathbf{E}) = V_k - V_3 = \Delta_k (\mathbf{0}) - e \ell_k |\mathbf{E}|, \quad
	\text{for } k =1, 2, 
\end{equation}
where $-e<0$ is electron charge and $\ell_{1(2)}$ stands for the distance between orbitals $1(2)$ and $3$ (see Fig.~\ref{fig:tight_binding}(b)). 
In the regime of low-energy physics, on-site repulsions $U_k$ are much larger than $V_k$ and $t_k$. For simplicity, we can assume the following: equal on-site energies for orbitals 1 and 2, i.e., $U_1 = U_2 = U$; zero on-site energy in orbital 3, $U_3 = 0$; and equal potential of empty orbitals, $\Delta_1 (\mathbf{0}) = \Delta_2 (\mathbf{0}) = \Delta$. Thus, up to second-order perturbation, we obtain the effective Heisenberg Hamiltonian
\begin{equation}
	H_\text{eff} = J (\mathbf{E}) \; \bm{S}_1 \cdot \bm{S}_2,
\end{equation}
where the exchange coupling, now a function of electric field, takes the following form
\begin{equation}
	J(\mathbf{E}) = \frac{2 \vert t_1 t_2 \vert^2}{(U + \Delta_1(\mathbf{E}))^2 (U + \Delta_2(\mathbf{E}))^2} \left[ 
	\frac{(U + \Delta_2 (\mathbf{E}))^2}{U + \Delta_1(\mathbf{E})-\Delta_2(\mathbf{E})} 
	+\frac{(U + \Delta_1 (\mathbf{E}))^2}{U - \Delta_1(\mathbf{E})+\Delta_2(\mathbf{E})} 
	+ 2U + \Delta_1(\mathbf{E}) + \Delta_2(\mathbf{E})
	\right].
\end{equation}
In the absence of electric field, the Heisenberg exchange coupling simply reads
\begin{equation}
	J_0 = J(\mathbf{0}) = \frac{4 \vert t_1 t_2 \vert^2}{(U + \Delta)^2} \left( \frac{1}{U} + \frac{1}{U + \Delta} \right).
	\label{eq:J_0}
\end{equation}

We confirmed that Eq.~(\ref{eq:J_0}) reproduces experimental values for $J_0$ for atomic dimers. In Fig.~\ref{fig:tight_binding}(c), we find that for $U = 3 \text{ eV}$ and $0.05<t<0.15 \text{ eV}$, the resulting $J_0$ falls between $0.1<J_0<30$ GHz, which is in line with the values found for Ti-Ti pairs on MgO at a distance range of $0.7<d<1.2$ nm~\cite{bae2018tipairs}. In the same experiment $J_0$ was found to exponentially depend on the distance $d$ between the Ti atoms as $J(d)=J_0\exp(-d/\lambda)$ with $\lambda = 64.6$ pm. Since $J \propto t^4$, we can convert the hopping into distance as shown in Fig.~\ref{fig:tight_binding}(d).

\begin{figure}
	\centering
	\includegraphics[width=0.7\linewidth]{ 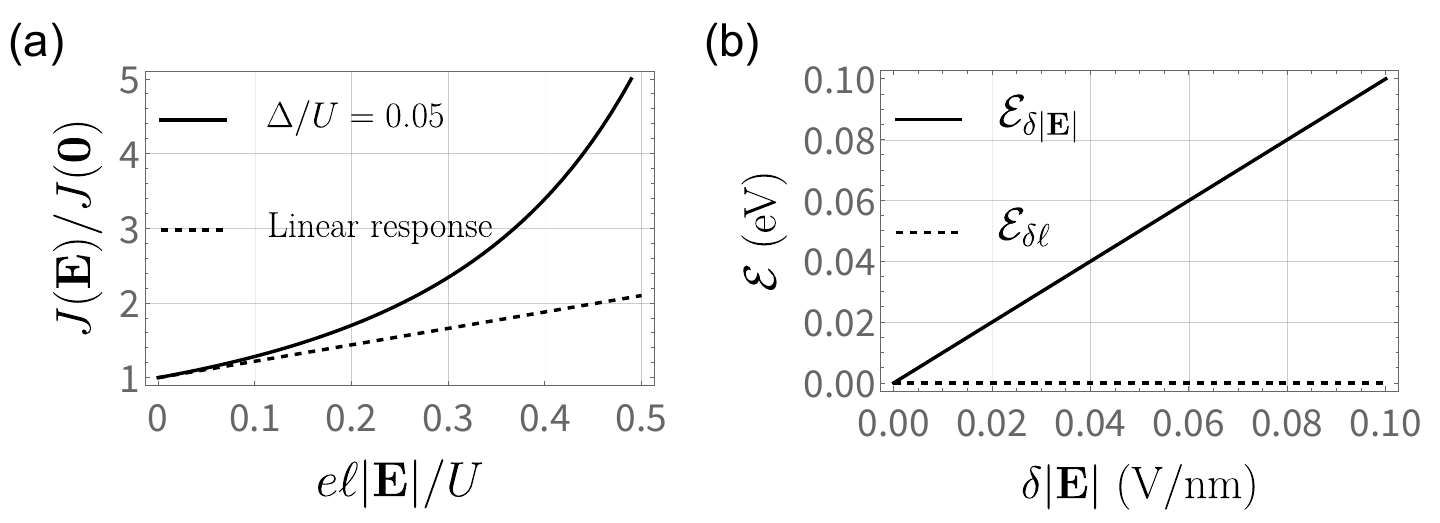}\caption{\textbf{Exchange coupling modulation:} (a) Plot of exchange coupling strength as a function of electric field strength. $J(E)$ is linear if $e\ell |\mathbf{E}| / U \lesssim 0.1$ (in case that $U_1 = U_2 = U$ and $U_3 = 0$). (b) The contribution of $\delta \mathbf{E}$ and $\delta \ell$ to the electric energy $\mathcal{E}$, computed with $|\mathbf{E}| = 0.1$~V/nm and $\ell= 1$~nm. The contribution of varying vertical distance is virtually zero. We assume the linear relation between vertical distance and electric field, with the slope taken from Table~\ref{tab:dft_displacement} below.}\label{fig:hopping_check}
\end{figure}

\subsection{Modulation of the superexchange by E-fields}

Under the assumption that the electric energy is much smaller than site energy difference and on-site repulsion
\begin{equation}
	e \ell_k |\mathbf{E}| \ll \Delta\ll U \; \text{ for }  k = 1, 2,
\end{equation}
we can expand the exchange coupling using Taylor series which up to the first order, is written as follows
\begin{equation}
	J(\mathbf{E)} = J_0 \left[ 1 + \frac{1}{2} 
	\left(\frac{3}{U + \Delta} - \frac{1}{2 U + \Delta} \right) e
	\ell \vert \mathbf{E} \vert \right].
\end{equation}
with $\ell \equiv \ell_1 + \ell_2$.
Thus, under influence of RF electric field $|\mathbf{E}| \propto V_\text{RF} \cos (\omega_{ij} t)$, the amplitude of modulated exchange coupling takes the following form
\begin{equation}
	J_1 = \frac{J_0}{2} 
	\left(\frac{3}{U + \Delta} - \frac{1}{2 U + \Delta}  \right) e
	\ell \vert \mathbf{E} \vert \propto J_0 V_\text{RF}. 
\end{equation}
Hence, we conclude that under electric field modulation, the spin coupling strength is modulated with a small magnitude and an identical frequency. The underlying mechanism is the electric field modulating on-site energy of Ti orbital, resulting in the modulated coupling strength.

We would like to further comment on the response of exchange coupling to the electric field, as depicted in Fig.~\ref{fig:hopping_check}(a). Since the lengths $\ell_{1 (2)}$ are in order of 0.1 nm and the electric field $| \mathbf{E} |$ is in order of 0.1 V/nm, the electric energy is very small compared to the typical on-site repulsion energy ($\sim 1 \text{ eV}$). Specifically, $e\ell |\mathbf{E}| /U \ll 1$ holds, ensuring the linear response regime. Further, we estimate that the dominant contribution to the potential variation stems from the AC modulation of the electric field but not the piezoelectric displacement. Denoted the electric energy as $\mathcal{E}$, it can be written as the contribution from modulation of electric field $\mathcal{E}_{\delta \mathbf{E}}$ and piezoelectric displacement $\mathcal{E}_{\delta \ell}$ as follows
\begin{equation}
	\delta \mathcal{E} \approx e \ell \delta |\mathbf{E}|  + e |\mathbf{E}| \delta \ell = \mathcal{E}_{\delta \mathbf{E}} + \mathcal{E}_{\delta \ell},
\end{equation}
where the displacement is computed by $\delta l = |a| \delta |\mathbf{E}|$ with the slope $a = -0.11 \times 10^{-3}$ nm$^2$/V;  taken from Table~\ref{tab:dft_displacement} below for a Ti-Fe pair and assuming the identical displacement for two atoms.
Fig.~\ref{fig:hopping_check}(b) shows that the change in length $\mathcal{E}_{\delta \ell}$ contributes much less than 1\% to the overall variation of the energy whilst moderate field strengths of 100 mV/nm lead to a significant change of the order of 0.1 eV.

\newpage
\section{Comparison of $J$-modulation magnitudes in various systems}
\label{sect:comparison}

{\corr{Modulation of the exchange coupling by electric fields is a common observation for several other systems, such as metallic thin films~\cite{Tsymbal2012Spintronics:Magnets, Mankovsky_prb_2021, Juarez-Reyes2014Electric-field-modulated1} and patterned ferromagnetic nanostructure~\cite{Matsukura2015ControlFieldsb}. Recently, it was proposed that a time-dependent gate voltage could modulate the exchange interaction  in molecular films~\cite{Dhingra_2022}.
		In Table~\ref{tab:exchange_coupling} we collect examples from literature for different systems and compare the achievable modulation $J_1$ per electric field strength and compare it with the modulation required to explain the Rabi rate of 10-20 MHz observed in the experiments on Fe-Ti pairs. We find that the ratio of observed modulation compared to the modulation in Fe-Ti atomic pairs can easily exceed 10 and can be as large as several thousand in molecular magnets. }}

\begin{table}[h!]
	\centering
	\begin{tabular}{|c|c|c|c|}
		\hline
		System & $J_1 \text{ (MHz)}$ @ $|\mathbf{E}| = 100 \text{ (mV/nm)}$ & Ratio & Ref. \\
		\hline 
		\hline
		Fe-Ti atomic pair & $20$  & 1 &~\cite{Bui2024All-electricalSpinb}  \\
		\hline
		\hline
		Fe-Fe atomic pair & $100$  & $5$ &~\cite{DFT_footnote} \\
		Fe-Ti atomic pair & $6$ & $0.3$ &~\cite{DFT_footnote} \\
		\hline
		\hline
		$\textrm{Cu}_3$ molecular magnet & $ (4 - 8.5) \times 10^2$ & $ 20 - 42.5 $ &~\cite{Islam_PRB2009_Cu3magnetic} \\
		Molecular magnet& $ (2.4 -12) \times 10^4 $ & $ 1200 - 6000$ &~\cite{Baadji_NatMat2009_electrostatic} \\
		Molecular magnet& $4.2 \times 10^4 $  & $2100$ &~\cite{Fittipaldi_NatMat2019_Electricfieldmodulation} \\
		Fe thin film& $60 - 121$  & $ 3 - 6$ &~\cite{Mankovsky_prb_2021} \\
		Co thin film& $2.41 \times 10^4$  & $1205$ &~\cite{Oba2015PRL, Ishibashi_2018} \\
		\hline
	\end{tabular}
	\caption{Reported exchange coupling modulation $J_1$ induced by electric field for various systems. $E$ denotes the change in the electric field amplitude. 
		{\corr{Ratio in the third column represents each row's value relative to the first row and for consistency, all values are represented using a fixed field $|\mathbf{E}| = 100 \text{ mV/nm}$.}}
		The first row corresponds to the experimentally available data.{\corr{ The second row are DFT calculations performed for this work.}} The first three rows are relevant to the system studied in this work.}
	\label{tab:exchange_coupling}
\end{table}

\newpage
\section{Spatial components of the exchange coupling modulation}
\label{sect:spatial_component}

\begin{figure}[h!]
	\centering
	\includegraphics[width=1\linewidth]{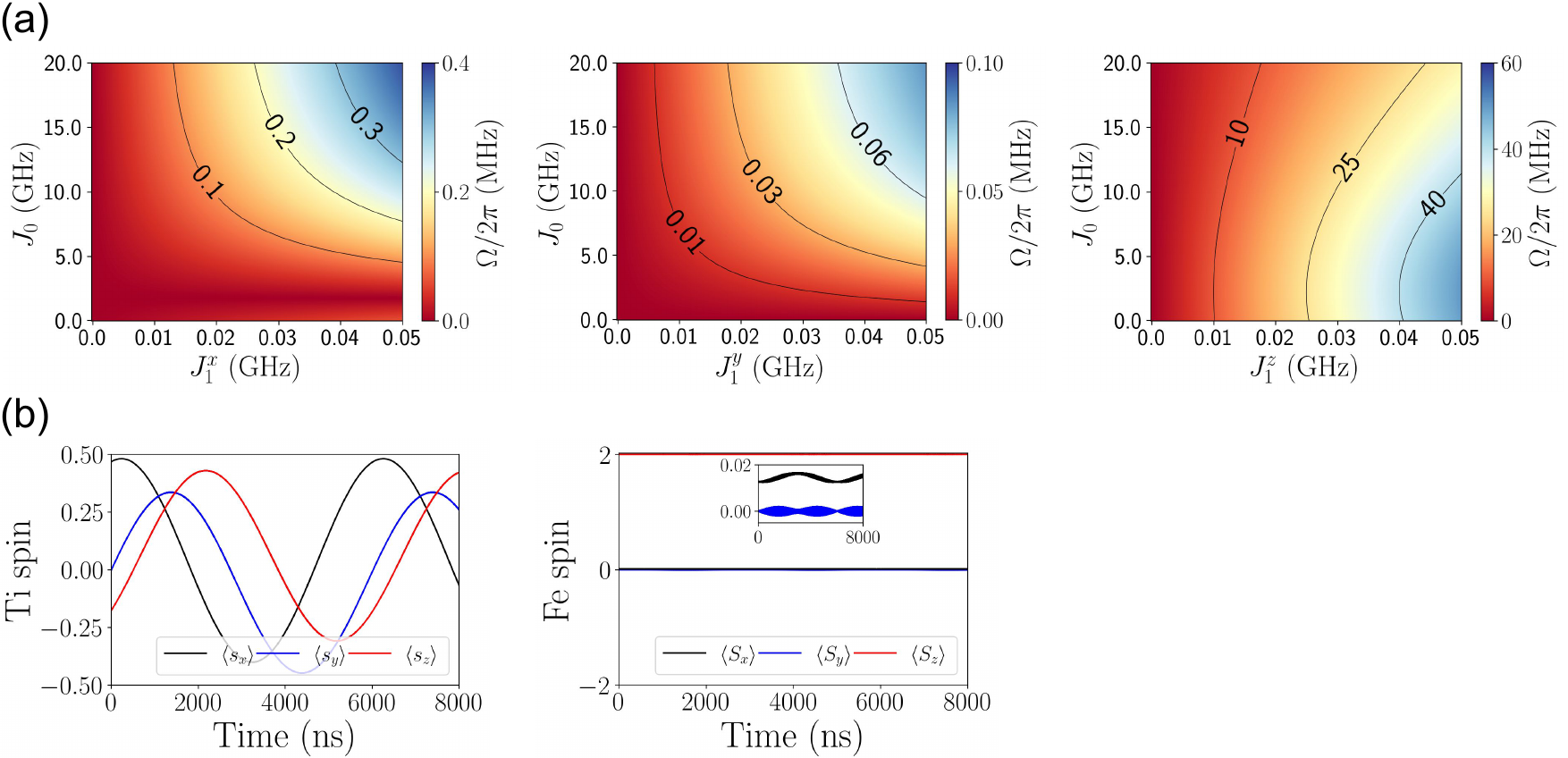}
	\caption{
		(a) Rabi rates computed by modulation of individual components of the exchange interaction. 
		(b) Spin dynamics of Ti-Fe pair using modulation of $x$-component exchange interaction with $J_0 = 6.8 \text{ GHz}, J_1^x = 0.05 \text{ GHz}, J_1^y = J_1^z = 0 \text{ GHz}$. The computed Rabi rate is $\Omega \approx 0.176 \text{ MHz}$.
		Fixed parameters are: $D_0 = -4.7 \text{ meV}$, $D_1 = 0 \text{ meV}$, $B_\text{ext} = 0.9 \text{ T}, \theta_\text{ext} = 80^\circ$. Isotropic $g$-tensor $g=2$.}
	\label{fig:Jxyz_modulation}
\end{figure}

In this section, we aim to analyze the contribution of each component of the exchange interaction to the Rabi rates. Instead of assuming isotropic modulation of $J_1$ as described in Eq. (3), we consider an independent modulation along each spatial direction, from each other, parameterized by $(J_1^x, J_1^y, J_1^z)$. Under this assumption, the Hamiltonian can be re-written as
\begin{equation}
	H_{J_1} (t) = (-J_1^x S_x s_x - J_1^y S_y s_y - J_1^z S_z s_z) \cos(\omega_{ij} t).
\end{equation}
The results for these independent modulations are shown in Fig.~\ref{fig:Jxyz_modulation}(a), where the modulation along $z$ direction dominates the other components. The $x$ and $y$-components of the exchange interaction just make tiny contributions to the Rabi rates. This is due to the specific out-of-plane anisotropy of the Fe, which forces the quantum magnet to align along the $z$-direction. Consequently, the direction of the anisotropy in a quantum magnet determines which components of the $J$-modulation will dominate over others. As a particular example for the system presented in this work, we choose $J_1^x = 0.05 \text{ GHz}$ to illustrate the spin dynamics. It results in a small Rabi rate, $\Omega \approx 0.176 \text{ MHz}$, with Rabi cycle lasting on the order of microseconds, see Fig.~\ref{fig:Jxyz_modulation}(b).

\newpage
\section{Detuning and phase control of the Rabi process}
\label{sect:detuning_phase}
\begin{figure}[h!]
	\centering
	\includegraphics[width=0.8\linewidth]{ 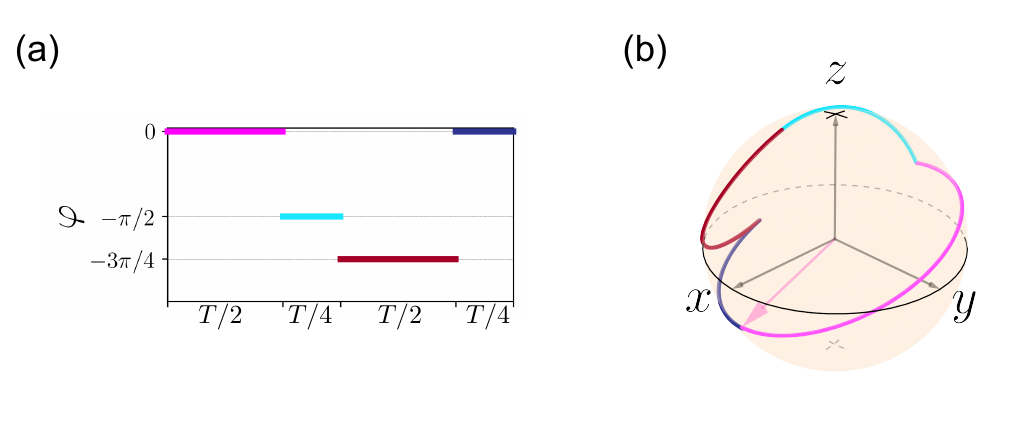}
	\caption{Effect phase control on the Rabi process. (a) Time-dependent driving phase $\varphi$ used for constructing a composite pulse sequence. Constant phases for a fixed time interval are employed, with $T$ being period of a Rabi cycle. (b) Spin evolution on the Bloch sphere where the color of each trajectory segment matches the phase color used in the corresponding time interval on the left. The parameters for the simulation are: $D_0 = -4.7 \text{ meV}$, $D_1 = 0.41 \text{ $\mu$eV}$, $J_0 = 6.8 \text{ GHz}$, $J_1 = 0.01 \text{ GHz}$, $B_\text{ext} = 0.9 \text{ T}, \theta_\text{ext} = 80^\circ$, and isotropic $g$-tensor $g=2$. }
	\label{fig:phase_control}
\end{figure}

\begin{figure}[h!]
	\centering
	\includegraphics[width=0.8\linewidth]{ 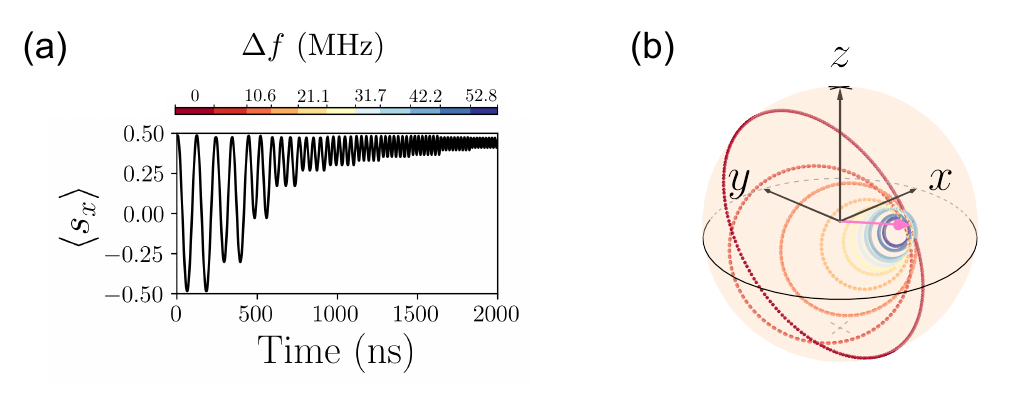}
	\caption{Effect of detuning on the Rabi process. (a) Time evolution of $\left< s_x \right>$ under frequency-varied driving. The colorbar represents the amplitude of the detuning frequency $\Delta f$. (b) Spin evolution on the Bloch sphere corresponding to each detuning amplitude. The parameters for the simulation are: $D_0 = -4.7 \text{ meV}$, $D_1 = 0.41 \text{ $\mu$eV}$, $J_0 = 6.8 \text{ GHz}$, $J_1 = 0.01 \text{ GHz}$, $B_\text{ext} = 0.9 \text{ T}, \theta_\text{ext} = 80^\circ$, and isotropic $g$-tensor $g=2$.}
	\label{fig:detuning}
\end{figure}

{\corr{To demonstrate that the driving resulting from $J$-modulation allows for full quantum state control of a qubit, we demonstrate two additional features: first, full quantum state control on the Bloch sphere is possible within our model by changing the phase $\varphi$ of the driving term $\cos ( \omega_{ij} t + \varphi )$, as shown in Fig.~\ref{fig:phase_control}. We intentionally choose constant phases over fix time intervals so that the spin evolution follows a piecewise-smooth trajectory. This simulation demonstrate arbitrary rotation on the Bloch sphere for the qubit. }
	The second feature is the detuning of the resonance frequency. It leads to the commonly observed increase in effective Rabi frequency by $\Omega'=\sqrt{\Omega^2+\Delta f^2}$, where the driving is on resonance when the detuning $\Delta f=0$, and the reduction of the amplitude of the coherent oscillations, rendering the increment in the Rabi rate useless for quantum control, which is a well known result. These properties can be seen at Fig.~\ref{fig:detuning}: after several full Rabi cycles, the driving frequency is stepwise increased, resulting in a reduced oscillation amplitude and a higher oscillation frequency.}

\section{Density functional theory calculations}
\label{sect:DFT_calculation}
We performed density functional theory (DFT) calculations as implemented in Quantum Espresso (V7.3)~\cite{Giannozzi_2009, Giannozzi_2017} which uses pseudopotentials and plane-waves to represent the wave functions. We used PAW pseudopotentials from the PSLibary~\cite{dalcorso2014} for all atom types and a cutoff for the kinetic energy of 90 Ry with a dual of 10 for the dense grid. Integration of the Brillouin zone was performed on a $2 \times 2 \times 1$ $k$-grid. The structure was build by first creating a 4 monolayers (ML) thick $6 \times 4$ slab of Ag (100) ($a=4.16$ \AA) capped by 2 ML of MgO. Then two Fe atoms (or one Fe and one Ti atom) were deposited two oxygen lattice sites apart ($d=8.320$ \AA) on the experimentally confirmed oxygen-top adsorption site and the whole system was relaxed until the total force was less than 0.01 Ry/a$_0$ (a$_0$ is the Bohr radius). During the initial relax calculations no electric field was applied. All cells were padded by 10 \AA~ of vacuum in $z$-direction.

\subsection{E-field dependent displacement}
\begin{figure}[h!]
	\centering
	\includegraphics[width=\linewidth]{ 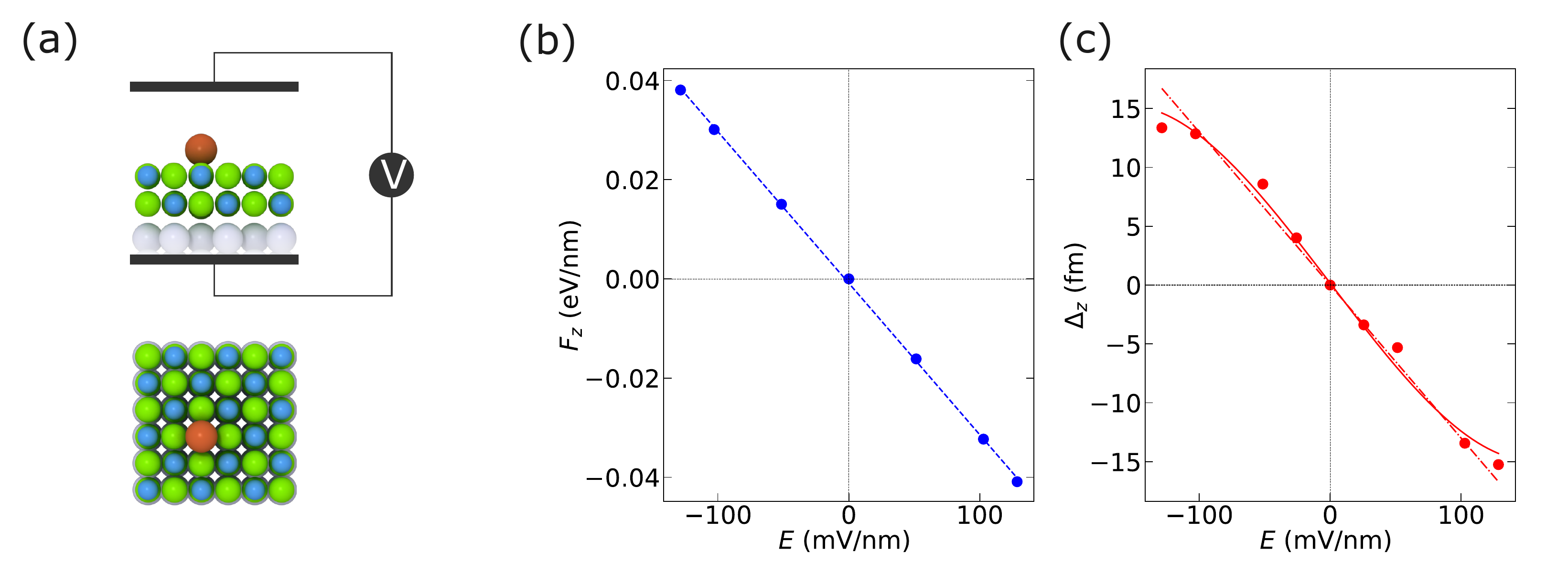}
	\caption{\textbf{E-field generated force and displacement:} (a) plate-capacitor model of the applied electric field for an Fe atom adsorbed on MgO/Ag. The silver is partially removed in the schematic for clarity. (b) Force induced on the Fe atom. The line is a linear fit. (c) Displacement of an Fe atom as function of the E-field. Overlayed are the fits with fitting results given in Tab.~\ref{tab:dft_displacement} } 
	\label{fig:fig_DFT_force}
\end{figure}

To estimated the displacement of the adatom under a perpendicular electric field we performed relax calculations of an Fe atom placed on an oxygen top site of an MgO/Ag slab with the bottom two layers of the silver frozen. Since surface adsorbed atoms have a non-vanishing net-charge and dipole due to the bonding geometry with the oxygen underneath the applied out-of-plane field generates a force. This field is applied in a plate capacitor model as shown in Fig.~\ref{fig:fig_DFT_force}(a) with a saw-like potential that is compatible with the periodic boundary conditions. The resulting force is shown in Fig.~\ref{fig:fig_DFT_force}(b) and exhibits a linear trend. The linear fits yield a slope of $\Delta F_z/\Delta V\approx -0.0008\pm0.0001$ eV/mV. If the system is allowed to relax (only the bottom two silver layers remained frozen) the resulting displacement is nonlinear as shown in Fig.~\ref{fig:fig_DFT_force}(c) where a second or third-order polynomial seems to give the best fit. Overall the displacement is very small (few fm) compared to the adsorption height of $1.79$ \AA. {\corr{An alternative perspective on this \textit{piezoelectric displacement} is that the ion positions are fixed but the charge of the atoms is modulated periodically, i.e. one could consider the Born effective charges of the crystal field~\cite{C7NR01596H}, which essentially leads to the same conclusion as displacement of the adatom.}}

The non-linearity is another factor that indicates that this \textit{piezoelectric displacement} is not the dominant driving factor as in recent experiments the Rabi rate has been shown to be strictly linear to the applied RF voltage~\cite{Phark2023Electric-Field-DrivenMagnet, Phark2023Double-ResonanceSurface, Bui2024All-electricalSpinb}. We further note that our displacements are smaller by a factor of about 10 compared to previous calculations that used 3 ML of MgO and no silver substrate~\cite{Seifert2020LongitudinalMicroscope}. This also means that values used in the main text to estimate the driving efficiency from $D$-modulation are most likely overestimating the driving efficiency yet still yield insufficiently large Rabi rates.

\begin{table}[h!]
	\centering
	\begin{tabular}{c|c|c|c}
		model& parameters & $R^2$\\
		\hline
		$a\cdot x$ &  $a=-0.11\pm0.005$ & 0.98\\
		$a\cdot x+b \cdot x^2$ & $a=-0.12\pm0.005$, $b=-3\cdot 10^{-5}\pm 2\cdot 10^{-5}$ & 0.98\\
		$a\cdot x+b \cdot x^3$ &$a=-0.15\pm0.01$, $b=2\cdot 10^{-6}\pm 9\cdot 10^{-7}$ &0.99\\
	\end{tabular}
	\caption{Fitting results for different fit functions of the DFT displacement shown in Fig.~\ref{fig:fig_DFT_force}(c); displacement in fm and field strength in mV/nm.}
	\label{tab:dft_displacement}
\end{table}

\subsection{E-field dependent exchange coupling}
\begin{figure}[h!]
	\centering
	\includegraphics[width=0.9\linewidth]{ 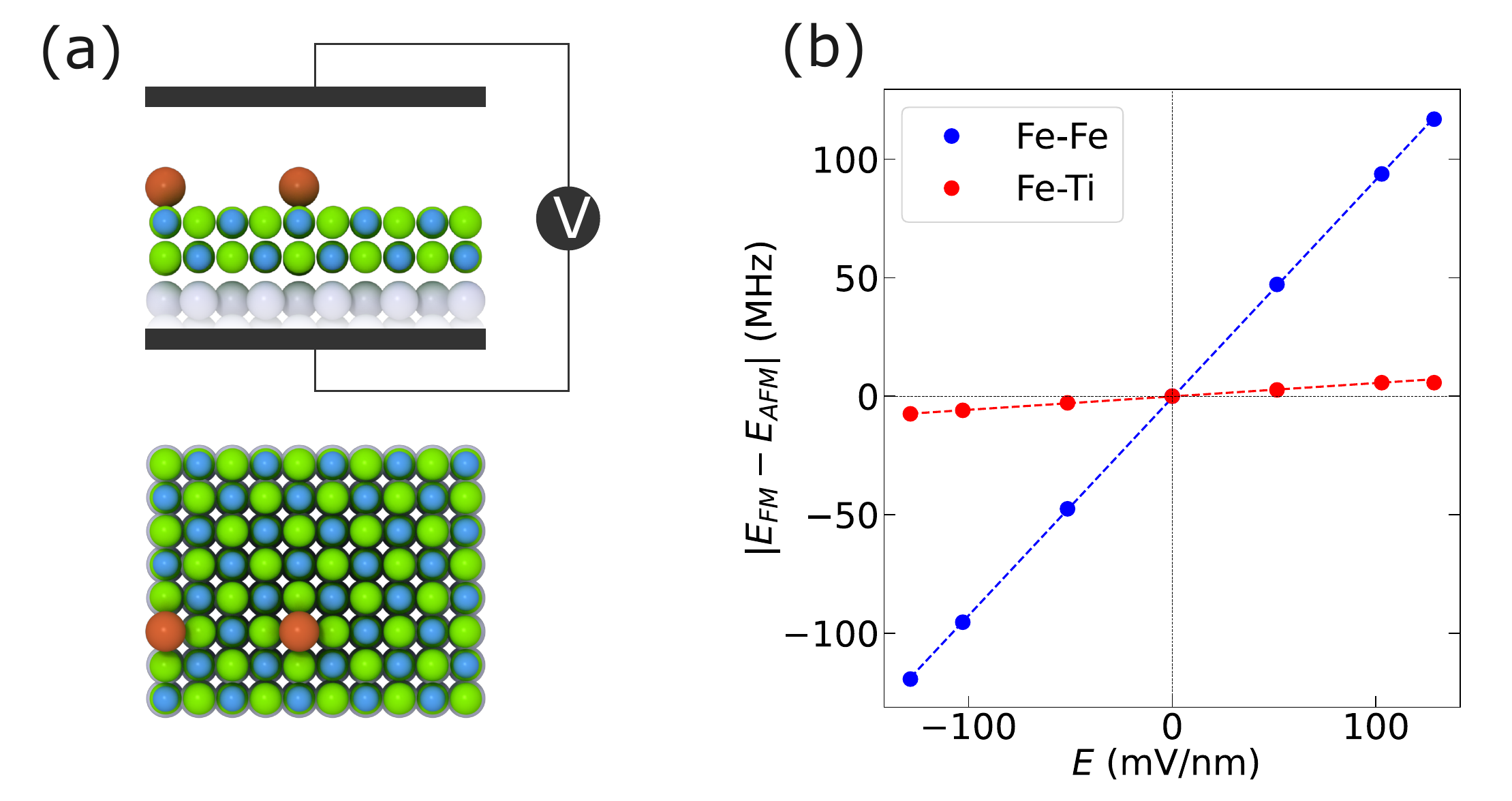}
	\caption{\textbf{E-field dependence of exchange coupling energy:} (a) Schematic of the DFT cell setup. Two Fe atoms with lateral separation $d=8.32$ \AA~in a plate-capacitor. (b) The resulting energy difference between FM and AFM coupled state as function of the electric field for an Fe-Fe (blue) and Fe-Ti (red) pair. The zero-field value of the exchange  coupling $E_0$ is subtracted in both cases. } 
	\label{fig:fig_DFT_exchange}
\end{figure}

To show that the electric-field modulation of the exchange coupling between two spin centers adsorbed on thin insulating layers deposited on a metal is a general phenomenon we calculated the electric field dependence of the exchange coupling constant of atomic pairs at fixed distance of $d=8.32$ \AA. We used DFT calculations with the same plate capacitor model as before with an out-of-plane field applied as shown in Fig.~\ref{fig:fig_DFT_exchange}(a). For a pair of two Fe atoms we calculated the total energy of the high spin ferromagnetic (FM,  $M=7.76~\mu_B$) coupled state as well as the low-spin antiferromagnetic (AFM, $M=0.0~\mu_B$) state. Their relative energy difference as function of the electric field is shown in Fig.~\ref{fig:fig_DFT_exchange}(b) with a linear fit yielding a slope of $E_1=-0.91\pm0.005$ MHz/(mV/nm) within the applied E-field range of $V \approx \pm 130$ mV/nm. We note that the FM coupled magnetic moment has a very slight dependence on the applied external field but stays within $M=7.67\pm0.1~\mu_B$. In all cases the pair is FM coupled. We note that this is in contrast to Fe-pairs measured on MgO/Ag at low external magnetic fields, at distances $>10$ \AA~ which were found to have AFM ground state attributed to the dipolar coupling mechanism~\cite{choi2017}. Deviations from the dipolar coupling regime for Fe-Fe pairs were found experimentally below 10 \AA~\cite{Choi2017a}. The situation is similar for Fe-Ti pairs (FM,  $M=5.3 ~\mu_B$, AFM,  $M=2.4~\mu_B$) at the same distance, however the exchange coupling energy $E_0$ as well as the slope of the field modulation $E_1$ is generally smaller in Fe-Ti compared to Fe-Fe (see Tab.~\ref{tab:exchange_coupling}).  

To estimate $J_0$ from the DFT total energies of FM and AFM state one could employ a number of different approaches\cite{Schurkus2020}. The simplest is the Noodleman formula~\cite{noodleman1981valence} which assigns $J$ based on the FM and AFM total energies $E$ as Eq.~(\ref{eq:noodlman})

\begin{equation}
	J_0=-\frac{E_{\rm FM}-E_{\rm AFM}}{s_{\mathrm{max}}^2}, 
	\label{eq:noodlman}
\end{equation}
with $s_{\mathrm{max}}$ being the maximal total spin of the two spin centers. In our case, for a separation of 8.32 \AA, we obtain $J_0\approx 100$ MHz for Fe-Fe pairs and $J_0\approx 50$ MHz for Fe-Ti pairs. The subscript in $J_0$ emphasizes that this is the static component at zero applied field. 

{\corr{We note that the pair distance in the simulations is slightly larger than the largest measured pair~\cite{Phark2023Electric-Field-DrivenMagnet}, which will reduce $J_0$ (and hence $J_1$). We further note that DFT calculations of exchange coupling constants suffer from accuracy issues due to the single-determinant nature of the wave function, which leads to spin-contamination, especially of the low-spin solution. Further, the exchange coupling constants can show some dependence on the exchange-correlation functional used, making predictive calculations difficult~\cite{Sheng2020}. A higher level of treatment for the exchange-correlation functional would be desirable, however the large size of the system including metal substrate, insulating layer and adatoms makes such a calculation prohibitively expensive. Finally, we note that we are only exploring the \textit{trends} of the exchange coupling as a function of the applied electric field, for which we believe DFT yields relatively reliable results, notwithstanding the points mentioned above.}}

\begin{table}[h!]
	\centering
	\begin{tabular}{c|c|c}
		System& $\left| E_0\right|$ (MHz) & $\left| E_1\right|$ (MHz/mV/nm)\\
		\hline
		Fe-Fe& 1534 & $9.1\times 10^{-1}$\\
		Ti-Fe& 330 & $5.8\times 10^{-2}$\\
	\end{tabular}
	\caption{Exchange coupling strength $E_0$ and exchange coupling modulation strength $E_1$ obtained from DFT calculation for Fe-Fe and Fe-Ti pairs, separated by 8.320 \AA~on 2 ML of MgO/Ag (001).}
	\label{tab:exchange_coupling2}
\end{table}

We emphasize that in all calculations for the exchange coupling modulation the atom positions are frozen and the system is not allowed to relax.

\end{document}